\documentstyle[aps,epsfig,amssymb,floats]{revtex}

\begin{document}
\draft

\twocolumn[\hsize\textwidth\columnwidth\hsize\csname
@twocolumnfalse\endcsname

\title{Wannier-function description of the electronic polarization and infrared
absorption of high-pressure hydrogen}

\author {Ivo Souza$^{1}$, Richard M. Martin$^{1}$, Nicola
    Marzari$^{2}$, Xinyuan Zhao$^{3}$, and David Vanderbilt$^{3}$}

\address{$^{1}$ Department of Physics and Materials Research
        Laboratory, University of Illinois, Urbana, 
        IL 61801 \\
     $^{2}$ Department of Chemistry, Princeton University,
        Princeton, NJ 08544-1009 \\
     $^{3}$ Department of Physics and Astronomy, Rutgers
        University, Piscataway, NJ 08854-0849
    }

\date{\today}

\maketitle

\begin{abstract}
We have constructed maximally-localized Wannier functions for prototype
structures of solid molecular hydrogen under pressure, starting from LDA and 
tight-binding Bloch wave functions. Each occupied Wannier function can be 
associated with two paired protons, defining a ``Wannier molecule''. The sum of
the dipole moments of these ``molecules'' always gives the correct 
macroscopic polarization, even under strong compression, when the overlap
between nearby Wannier functions becomes significant. We find that at megabar 
pressures the contributions to the dipoles arising from the overlapping tails 
of the Wannier functions is very large. The strong vibron infrared absorption 
experimentally observed in phase III, above $\sim$150~GPa, is analyzed in terms
of the vibron-induced fluctuations of the Wannier dipoles. We decompose these 
fluctuations into ``static'' and ``dynamical'' contributions, and find that at
such high densities the latter term, which increases much more steeply with 
pressure, is dominant.
\end{abstract}


\vskip2pc]

\columnseprule 0pt

\narrowtext

\section{INTRODUCTION}
\label{intro}

\subsection{Wannier functions}
\label{intro_wf}

The electronic structure of periodic solids is usually described, in the
independent-electron approximation, in terms of the extended Bloch 
eigenfunctions. An alternative representation is provided by the Wannier 
functions (WFs)\cite{wannier37,blount}, which are localized, with a typical 
spread of the order of the atomic dimensions; 
they can be obtained via a 
unitary transformation of the Bloch states belonging to an isolated 
band\cite{wannier37} or to a composite group of bands\cite{blount,marzari97} 
(i.e., bands that may be connected among themselves by degeneracies,
but are separated from all others by energy gaps).
For some purposes the latter description is advantageous: for instance, the WFs
constructed from the states in the valence bands provide an intuitive, 
``chemical-like'' localized picture of bonding and dielectric properties of 
insulators\cite{marzari97}.

The major drawback of the Wannier representation is the strong nonuniqueness of
the WFs: their average location, shape, and spread 
all depend on the arbitrary choice of gauge\cite{blount,marzari97}.
In practice, this indeterminacy can be resolved by working with the set of 
WFs which is most localized according to some sensible criterion. A certain
degree of
arbitrariness still remains regarding which measure of localization to use, and
in fact several alternatives have been proposed in the literature.
We follow the approach of Ref. \cite{marzari97}, which amounts to
minimizing the sum of the quadratic spreads of the WFs (see Sec.~\ref{maxloc}).
We will use density-functional theory in the local density 
approximation (LDA), complemented by a tight-binding analysis, to investigate
in terms of well-localized WFs the electronic structure and dielectric
properties of compressed molecular
hydrogen\cite{foot-andreoni}.

\subsection{Compressed molecular hydrogen}
\label{intro_h}

Solid hydrogen under pressure has attracted considerable attention over the 
years\cite{mao94}, and the main goal has been to try to metallize it; this is 
expected to occur at high enough pressures, either by band gap closure in a 
molecular phase, or by molecular dissociation, whichever occurs 
first\cite{johnson00,stadele00}. However, up to 
the highest pressures reached so far ($\sim 340$~GPa), hydrogen appears to 
remain both molecular and insulating\cite{narayana98}. Nevertheless, a rich 
phase diagram has emerged, with three distinct phases unambiguously identified
using Raman and infrared (IR) spectroscopy\cite{mao94}.

\begin{figure}
\centerline{\epsfig{file=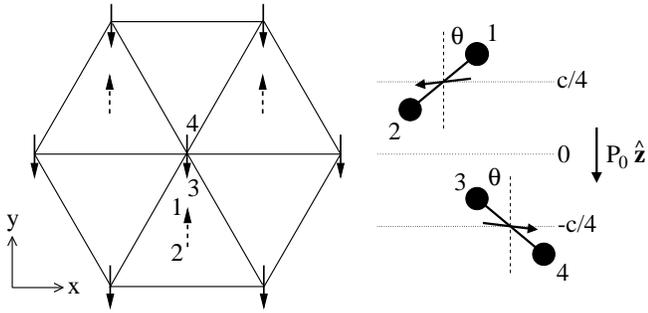,width=3.35in}}
\vspace{0.2cm}
\caption{
The $Cmc2_1$ structure viewed along the c-axis (left) and in the
$yz$ plane (right). The centers of the molecules lie on hcp sites, and
the molecules in the two sublattices are tilted away from the c-axis
by opposite angles $\theta$ and $-\theta$. The $C2/m$ structure is identical
except that the two molecules in the primitive cell are tilted in the same 
direction by an angle $\theta$. The arrows on the right side indicate the 
directions of the dipoles of the two ``Wannier molecules'' for $r_s=1.52$.
}
\label{fig_cmc21}
\end{figure}

The precise crystal structure of the high-pressure phases (phases II and III) 
has not been determined experimentally, and conclusive theoretical predictions
have proven quite difficult, due to the quantum effects
associated with the protons.
The purpose of the present work is not to propose new candidate structures, but
rather to make some very general points, illustrated on a couple of 
particularly simple prototype structures 
(Fig. \ref{fig_cmc21}), which were chosen mainly for 
clarity. It is hoped that, even if none of them turns out to be the correct 
structure of phase III (which is likely to be the case), they manage to capture
some of its 
relevant features.  For instance, {\it ab-initio} calculations at megabar 
pressures and low temperatures  tend to favor structures in which the centers 
of the molecules form an hcp or, more generally, a triangular 
lattice\cite{kaxiras92,kohanoff99} (possibly with a
small distortion\cite{edwards97})\cite{foot-cmca}.
It is believed that in phase III the molecules are 
orientationally ordered, with their 
axes tilted away from the $c$-axis, as such canted structures
tend to be more stable\cite{stadele00,kaxiras92,kohanoff99,kitamura00}. 
Moreover, the resulting lowering of symmetry 
gives rise to IR-active vibron modes\cite{zallen94,cui95}; indeed, one of the 
signatures of phase III, above 150 GPa,
is a strong IR absorption peak in the vibron frequency 
range\cite{hanfland93,hemley97},
which contrasts with the much weaker absorption found in phases I and II. 

\subsection{Organization}\label{organization}

The paper is organized as follows.  In Sec.~\ref{maxloc} we briefly review the 
method used for constructing well-localized WFs; in Sec.~\ref{eq_dipole} we 
investigate the permanent dipole moments of the ``Wannier molecules'' as a 
function of crystal structure and pressure; the results are presented in terms
of a ``static effective charge'' vector associated with each molecule. 
In Sec.~\ref{dipole_fluct} we look at the vibron-induced fluctuations of 
those dipoles, which can be quantified in terms of a ``vibron-induced effective
charge'' vector on each molecule. From the effective charges the strength of 
the vibron IR absorption is calculated and compared with experiment, and the 
relative importance of the ``static'' and ``dynamical'' charge contributions 
is ascertained. Sec.~\ref{extent} deals with the spatial distribution of the 
WFs and the effect of molecular overlap on the dielectric properties of the 
compressed solid.
In Sec.~\ref{uniqueness} we investigate the nonuniqueness associated with the
definition of well-localized WFs in the dense solid. The tight-binding analysis
is presented in Sec.~\ref{tb}. 
In Sec.~\ref{discussion} we give a discussion of our results.

Atomic units are used for all quantities except pressures, which are in
gigapascal (GPa), and energies, in electron-volt (eV).
Densities are expressed in terms of $r_s$, defined as
$(4\pi r_s^3/3)a_0^3=V/N$, where $N$ is the number of protons in the volume
$V$, and $a_0$ is the Bohr radius. Since the LDA tends to underestimate the 
pressures, in order to convert from $r_s$ to pressure we use an 
experimental equation of state extrapolated to high 
pressures\cite{loubeyre96}.
The LDA calculations were performed using a plane-wave 
cutoff of $90$~Ry and
the bare Coulomb potential of the protons. The self-consistent calculations
with 4 atoms per cell used a $(11,11,11)$ Monkhorst-Pack mesh for the
Brillouin zone sampling. 
After self-consistency was achieved, the ``maxloc'' WFs were determined 
starting from the Bloch states again calculated on a $(11,11,11)$ mesh.
In all the calculations we have used the following parameters for the
$Cmc2_1$ structure described in Fig.~\ref{fig_cmc21}:
$r_{\rm bond}=1.445$~a.u., $c/a=1.576$, and the tilt angle
$\theta=54.0^\circ$. For $C2/m$ the parameters are
$r_{\rm bond}=1.456$~a.u., $c/a=1.588$, and $\theta=69.5^\circ$.
In both cases the structures were obtained by minimizing the enthalpy at a fixed
LDA pressure of $100$~GPa, with a resulting density of
$r_s=1.52$ (which experimentally corresponds to about $115$~GPa, according to 
Ref.~\cite{loubeyre96}). The same parameters were used at all other densities.

\section{Maximally-localized Wannier functions}\label{maxloc}

A set of WFs $\{ w_n({\bf r}-{\bf R}) \}$, each labeled by a
different Bravais lattice vector ${\bf R}$, can be constructed from the
Bloch eigenstates $\{ \psi_{n \bf k} \}$ in band $n$ using the unitary 
transformation

\begin{equation}
\label{wf}
w_n({\bf r}-{\bf R}) = \frac{v}{8 \pi^3} \int_{\rm BZ}
e^{- i {\bf k} \cdot {\bf R}} \psi_{n \bf k} d{\bf k},
\end{equation}

\noindent where $v$ is the volume of the unit cell of the crystal and
the integral is over the Brillouin zone.
Except for the constraint $\psi_{n,{\bf k}+{\bf G}} = \psi_{n \bf k}$
for all reciprocal lattice vectors ${\bf G}$,
the overall phases of the Bloch functions
$\psi_{n \bf k} = e^{i {\bf k} \cdot {\bf r}} u_{n \bf k}$ are at our 
disposal. However, a different choice of phases (or ``gauge''),

\begin{equation}
\label{gauget_isol}
u_{n \bf k} \rightarrow e^{i \varphi_n({\bf k})} u_{n \bf k},
\end{equation}

\noindent does not translate into a simple change of the overall phases of the
WFs; their shape and spatial extent will in general be affected,
while the location of their centers of charge remains invariant modulo a
lattice vector ${\bf R}$\cite{marzari97}.
If the band is isolated, Eq. \ref{gauget_isol} is the only allowed type of
gauge transformation for changing the WF $w_n({\bf r})$ associated with that 
band. In the case of a composite group of bands, the allowed
transformations are of the more general form

\begin{equation}
\label{gauget_composite}
u_{n \bf k} \rightarrow \sum_m U_{mn}^{(\bf k)} u_{m \bf k},
\end{equation}

\noindent where $U_{mn}^{(\bf k)}$ is a unitary matrix that mixes the bands at
every wave vector ${\bf k}$. Under this transformation the individual Wannier 
centers can shift, but their sum over the group of bands is preserved modulo a
lattice vector~\cite{marzari97}.
Once a measure of localization has been chosen and the group of bands
specified, the search for the corresponding set of 
``maximally-localized'' WFs becomes a problem of functional minimization
in the space of the matrices $U_{mn}^{(\bf k)}$.
The strategy of Ref. \cite{marzari97} is to minimize
the sum of the quadratic spreads of the Wannier probability distributions
$\{ {| w_n({\bf r}) |}^2 \}$, given by

\begin{equation}
\label{omega}
\Omega = \sum_n \left( {\left< r^2 \right>}_n - 
{\left< {\bf r} \right>}_n^2 \right),
\end{equation}

\noindent where the sum is over the chosen group of bands
(in our application they will be the valence bands), and
${\left< {\bf r} \right>}_n = \int {\bf r} {| w_n({\bf r}) |}^2 d{\bf r}$, etc.
Interestingly,
the resulting ``maximally-localized'' (or ``maxloc'') WFs turn out to be
real, apart from an arbitrary overall phase\cite{marzari97}. 

In numerical calculations 
the Bloch states $\psi_{n \bf k}$ are computed on a regular
mesh of $k$-points in the Brillouin zone;
the integral in Eq. \ref{wf} is then replaced by a sum over the points
in the mesh. In Ref. \cite{marzari97} an expression was derived for the 
gradient of the spread functional $\Omega$ with respect to an
infinitesimal rotation $\delta U_{mn}^{({\bf k})}$ of the set of Bloch 
orbitals, in terms of the Bloch functions in such a mesh.
The only information needed for calculating the gradient
are the overlaps $\left< u_{m \bf k} | u_{n,{\bf k}+{\bf b}} \right>$,
where ${\bf b}$ are vectors connecting each mesh point to its
near neighbors. Once the gradient is computed, the minimization can then
proceed via a steepest-descent or conjugate-gradient algorithm.

Since the Bloch eigenstates at different $k$-points are initially 
computed by independent numerical matrix diagonalizations, their 
phases are unrelated. As a consequence, the WFs obtained directly
from them using the discretized version of Eq. \ref{wf}
will be poorly localized, or not localized at all.
In practice the following strategy is used for preparing a better set of
Bloch states as the starting point for the minimization: one
chooses a set of localized ``trial functions'' in the unit cell, which 
constitute a rough initial
guess at the WFs; for solid hydrogen we use Gaussians 
on the centers of the molecules.
Then, a unitary rotation among the initial
Bloch orbitals is made in order to maximize their projections onto these
trial functions (the detailed procedure is described in Eqs. 62-64 of
Ref. \cite{marzari97}).
For a reasonable choice of the width of the Gaussians
(we have used a r.m.s. width of 1 \AA), the 
resulting rotated WFs are already extremely close to the 
``maxloc'' ones, as discussed in Secs.~\ref{uniqueness} and \ref{tb_formalism}.

\section{Equilibrium Wannier dipoles}
\label{eq_dipole}

The neutral entity composed of two
paired nuclei and the occupied ``maxloc'' Wannier orbital centered around them
forms a ``Wannier molecule'' in the bulk of the solid. In the low-density
limit the ``maxloc''  WFs become nonoverlapping and coincide with the ground 
state bonding orbitals of isolated ${\rm H}_2$ molecules; however, at the high
pressures we are interested in,
there is an appreciable overlap between neighboring WFs.
In what follows we will sometimes
loosely refer to the ``maxloc'' Wannier molecules in the dense solid simply as
``molecules''. One should keep in mind, however, that had we chosen a
different measure of localization, the resulting ``maximally-localized'' WFs
in the dense system would in general differ somewhat from the ones we obtain.
This nonuniqueness is intrinsic to WFs, and can be viewed as a manifestation of
the ambiguity that always arises when trying to define ``molecules'' in a
dense medium, either in terms of WFs or by other means. 
These issues will be discussed in Secs.~\ref{uniqueness} and
\ref{mol_in_solid}.

\begin{figure}
\centerline{\epsfig{file=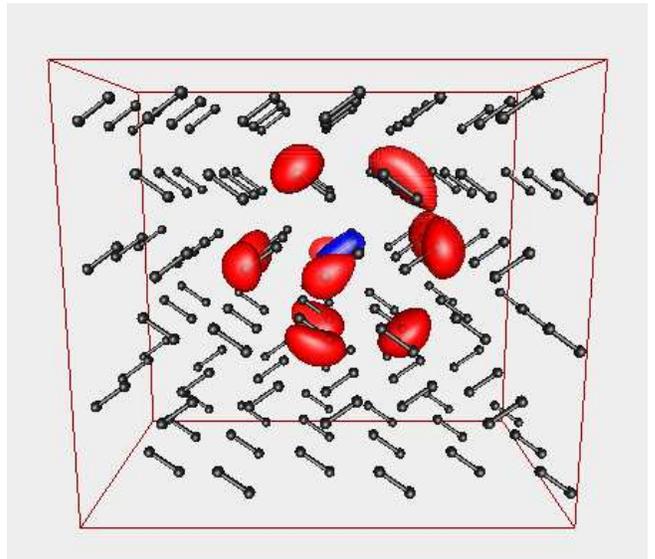,width=3.35in}}
\vspace{0.2cm}
\caption{
Contour plot of $\sqrt{v}w_1({\bf r})$ for
an occupied Wannier function in the $Cmc2_1$ structure at
$r_s=1.52$ ($v=59.3$~a.u.\ is the volume of the primitive cell).
The central, cylindrically-shaped contour, which represents the 
bonding part of the WF, has a positive amplitude of $+2.12$; 
the outer
lobes (``orthogonality tails''), with antibonding character, have an amplitude
of $-0.11$.
}
\label{fig_contour_eq_cmc21}
\end{figure}

Fig. \ref{fig_contour_eq_cmc21} shows a contour plot of a Wannier orbital for 
$Cmc2_1$ at $r_s=1.52$. The central positive contour with
a large amplitude represents the molecular bond. The lowering of 
symmetry due to the crystalline environment is clear from the shape of the 
outer ``corona'' formed by the negative lobes, which have an antibonding 
character.
These so-called ``orthogonality tails'' appear when the molecules overlap, due 
to the orthogonality requirement between different WFs;
they are concentrated around the twelve nearest 
molecules, 
which allows for an efficient orthogonalization between neighboring WFs.
We will argue in Sec.~\ref{extent} that these overlapping orthogonality tails
strongly influence the dielectric properties, and in particular the vibron
IR activity.

\begin{table}
\caption{Static (Eq. \ref{static_charge}) and 
vibron-induced (Eq. \ref{vibron_charge}) effective charge vectors for the two 
molecules in the primitive cell of the $Cmc2_1$ structure. The $x$-components 
vanish by symmetry.}
\begin{tabular}{ccccccc}
Molecule & \multicolumn{2}{c}{Static} & \multicolumn{2}{c}{In-phase} & 
\multicolumn{2}{c}{Out-of-phase} \\
\tableline
\mbox{1} & $q_y^{\rm s}(1)$ & $q_z^{\rm s}(1)$ &
$q_y^{\rm i}(1)$ & $q_z^{\rm i}(1)$ & 
$q_y^{\rm o}(1)$ & $q_z^{\rm o}(1)$\\
\mbox{2} & $-q_y^{\rm s}(1)$ & $q_z^{\rm s}(1)$ &
$-q_y^{\rm i}(1)$ & $q_z^{\rm i}(1)$ & 
$q_y^{\rm o}(1)$ & $-q_z^{\rm o}(1)$
\end{tabular}
\label{table_eff_charges_cmc21}
\end{table}

\begin{table}
\caption{Same as Table~\ref{table_eff_charges_cmc21}, but for the
$C2/m$ structure.}
\begin{tabular}{ccccccc}
Molecule & \multicolumn{2}{c}{Static} & \multicolumn{2}{c}{In-phase} & 
\multicolumn{2}{c}{Out-of-phase} \\
\tableline
\mbox{1} & $q_y^{\rm s}(1)$ & $q_z^{\rm s}(1)$ &
$q_y^{\rm i}(1)$ & $q_z^{\rm i}(1)$ & 
$q_y^{\rm o}(1)$ & $q_z^{\rm o}(1)$\\
\mbox{2} & $-q_y^{\rm s}(1)$ & $-q_z^{\rm s}(1)$ &
$-q_y^{\rm i}(1)$ & $-q_z^{\rm i}(1)$ & 
$q_y^{\rm o}(1)$ & $q_z^{\rm o}(1)$
\end{tabular}
\label{table_eff_charges_c2m}
\end{table}

An important effect of the anisotropic crystalline environment is
that the molecules becomes polarized under the
self-consistent internal electric fields inside the solid. In commonly used 
treatments of the dielectric properties of molecular 
crystals\cite{califano,munn88}, the so-called Clausius-Mossotti approximation 
is assumed: the system is modeled as a sum of {\it nonoverlapping} molecular
charge distributions which become polarized in the local field produced by the
surrounding molecules; the bulk polarization is then the sum of the individual,
nonoverlapping, molecular dipole moments, which can be straightforwardly calculated
from the bulk charge density $\rho({\bf r})$. Such a description becomes 
inappropriate whenever there is significant molecular overlap\cite{smith71}:
the electron density becomes different from zero everywhere, and as a result
the net dipole moment becomes dependent on the particular choice made for the unit
cell\cite{rmm74,ksv}. In the case of molecular crystals, this is expected to 
occur, for instance, when the system is strongly 
compressed\cite{foot-molecular-crystals}; under such circumstances a more 
careful treatment of the macroscopic polarization is required. According to the
Berry phase theory of bulk polarization\cite{ksv}, whenever such
overlap effects are significant, the macroscopic polarization 
${\bf P}_{\rm mac}$ of an insulating
crystal cannot be extracted from the bulk $\rho({\bf r})$, and is instead 
given, in the independent-electron approximation, by a Berry phase of the 
occupied Bloch states. This is a gauge-invariant quantity, and it
is identical to another invariant, 
the total sum of the dipoles ${\bf d}(n)$
of the ``Wannier molecules''\cite{ksv}:  
\begin{equation}
\label{bulk_P}
{\bf P}_{\rm mac} = \frac{1}{v} \sum_{n=1}^M {\bf d}(n), 
\end{equation}
\begin{equation}
\label{wf_dipole}
{\bf d}(n) = -2e\int {\bf r} \, {\left| w_n({\bf r}) \right|}^2\,d{\bf r},
\end{equation}
where $M$ is the number of valence bands
and $e$ is the magnitude of the electron 
charge. In Eq. \ref{wf_dipole} the origin is chosen 
midway between the two paired protons, to cancel their contributions.
The factor of 2 comes from spin-degeneracy (each occupied WF carries two
electrons).
We stress that Eq. \ref{bulk_P} does not rely on the Clausius-Mossotti 
approximation at all, and it remains exact even when the 
``Wannier molecules'' overlap strongly, as long as the system remains 
insulating.
The decomposition of ${\bf P}_{\rm mac}$ into individual Wannier dipoles is a 
powerful analysis tool, allowing us to go beyond the Berry phase approach used
in Ref. \cite{souza98}, which only gives the {\it net} polarization of the unit
cell.

\widetext
\begin{table*}
\caption[]{Static (Eq. \ref{static_charge}) and 
vibron-induced (Eq. \ref{vibron_charge}) effective charge vectors 
for molecule 1, for the hcp-centered (on-site) 
and the off-site structures, at $r_s=1.52$. Results are presented for the 
Wannier functions in the LDA approximation (WF) and for the electric quadrupole
model (EQ). The $x$-components vanish by symmetry.
$q_{\parallel}$ and $q_{\perp}$ are the magnitudes of the projections along the
molecular axis and perpendicularly to it, respectively.
}
\begin{tabular}{cc|dddd|dddd|dddd}
\multicolumn{2}{c}{} & \multicolumn{4}{c}{\mbox{Static}} & 
\multicolumn{4}{c}{\mbox{In-phase vibron}} & 
\multicolumn{4}{c}{\mbox{Out-of-phase vibron}} \\
\multicolumn{1}{c}{\mbox{Structure}} & \multicolumn{1}{c}{\mbox{Model}} & 
\multicolumn{1}{c}{\mbox{$q^{\rm s}_y(1)$}} & 
\multicolumn{1}{c}{\mbox{$q^{\rm s}_z(1)$}} & 
\multicolumn{1}{c}{\mbox{$q^{\rm s}_{\parallel}$}} & 
\multicolumn{1}{c}{\mbox{$q^{\rm s}_{\perp}$}} &
\multicolumn{1}{c}{\mbox{$q^{\rm i}_y(1)$}} & 
\multicolumn{1}{c}{\mbox{$q^{\rm i}_z(1)$}} & 
\multicolumn{1}{c}{\mbox{$q^{\rm i}_{\parallel}$}} & 
\multicolumn{1}{c}{\mbox{$q^{\rm i}_{\perp}$}} &
\multicolumn{1}{c}{\mbox{$q^{\rm o}_y(1)$}} & 
\multicolumn{1}{c}{\mbox{$q^{\rm o}_z(1)$}} & 
\multicolumn{1}{c}{\mbox{$q^{\rm o}_{\parallel}$}} & 
\multicolumn{1}{c}{\mbox{$q^{\rm o}_{\perp}$}}\\
\tableline
\mbox{$Cmc2_1$} & \mbox{WF} & $-$0.044 & $-$0.003 & 0.037 & 0.023 & 
$-$0.172 & $-$0.037 & 0.161 & 0.071 & 
$-$0.138 & $-$0.030 & 0.129 & 0.057\\ 
\mbox{(on-site)} & \mbox{EQ} & $-$0.014 &  $-$0.002 & 0.012 & 0.006 & 
$-$0.041 & $-$0.009 & 0.038 & 0.017 &
0.004 & $-$0.002 & 0.002 & 0.004\\ 
\tableline
\mbox{$Cmc2_1$} & \mbox{WF} & $-$0.074 &  $-$0.026 & 0.076 & 0.019 & 
$-$0.267 & $-$0.155 & 0.308 & 0.018 &
$-$0.281 & $-$0.043 & 0.258 & 0.119\\
\mbox{(off-site)} & \mbox{EQ} & $-$0.017 & $-$0.007 & 0.018 & 0.003 & 
$-$0.050 & $-$0.023 & 0.054 & 0.009 &
0.004 & 0.000 & 0.003 & 0.002\\
\tableline
\mbox{$C2/m$} & \mbox{WF} & $-$0.073 & $-$0.013 & 0.073 & 0.013 & 
$-$0.338 & $-$0.078 & 0.344 & 0.045 &
$-$0.437 & $-$0.086 & 0.439 & 0.073\\
\mbox{(on-site)} & \mbox{EQ} & $-$0.019 & $-$0.002 & 0.019 & 0.005 & 
$-$0.059 & $-$0.008 & 0.058 & 0.013 & 
0.004 & $-$0.002 & 0.003 & 0.003\\
\tableline
\mbox{$C2/m$} & \mbox{WF} & $-$0.118 & $-$0.008 & 0.114 & 0.032 & 
$-$0.500 & $-$0.054 & 0.489 & 0.115 &
$-$0.776 & $-$0.174 & 0.790 & 0.094\\
\mbox{(off-site)} & \mbox{EQ} & $-$0.023 & 0.002 & 0.021 & 0.009 & 
$-$0.068 & 0.001 & 0.064 & 0.023 & 0.005 & $-$0.004 &
0.003 & 0.006\\  
\end{tabular}
\label{table_eff_charges}
\end{table*}
\narrowtext

The permanent Wannier dipoles can be used to assign to each molecular 
charge distribution a ``static effective charge'' vector:
 
\begin{equation}
\label{static_charge}
{\bf q}^{\rm s}(n)=\frac{{\bf d}(n)}{r_{\rm bond}(n)}, 
\end{equation}

\noindent where $r_{\rm bond}(n)$ is the equilibrium bond length. This 
quantity, 
which vanishes in the low density limit of isolated molecules, measures the 
spontaneous symmetry-breaking charge transfer which occurs in the compressed 
solid whenever the two atoms in a molecule occupy crystallographically 
inequivalent sites. We emphasize that this 
definition is somewhat arbitrary, for the reasons discussed at the beginning of
this section
(but see Sec.~\ref{uniqueness}), and therefore
${\bf q}^{\rm s}(n)$
does not relate directly to any measurable quantity. However, it is a sensible
definition, which reduces to the natural one in the extreme ionic limit 
where the electron distribution is strongly concentrated around the ions.
In the next section we will use it to decompose 
the vibron effective charge into ``static'' and ``dynamical'' contributions, 
with the aim of understanding the origin of the strong IR absorption 
in phase III.

The location of the centers of the ``maxloc'' WFs reflects the 
symmetry properties of the crystal; this is apparent from
the form of the vectors ${\bf q}^{\rm s}(n)$, shown in the first set of
columns in
Tables \ref{table_eff_charges_cmc21} and \ref{table_eff_charges_c2m} for the 
two structures studied. The first set of columns in
Table \ref{table_eff_charges} lists their explicit values
for $r_s=1.52$, both with the molecules on-site and after allowing them to
move away from the ideal hcp sites\cite{edwards97,souza98}.
For comparison we also report the values calculated 
using an electric quadrupolar (EQ) model\cite{eq-model}.
In the $C2/m$ structure there is a center of inversion between the two 
molecules in the primitive cell, leading to a cancellation of their permanent 
dipoles; 
in the lower symmetry $Cmc2_1$ the $y$-components of the
individual dipoles still average to zero over the primitive cell, but the
$z$-components add up, yielding a small spontaneous polarization along the
$c$ axis (see Fig. \ref{fig_cmc21}).
In general $|q^{\rm s}_y| >> |q^{\rm s}_z|$,
so that the dipole moments
make an angle with the molecular axes, i.e.,
$q^{\rm s}_{\perp}$ in Table \ref{table_eff_charges} is nonzero, although 
it is smaller than $q^{\rm s}_{\parallel}$.
This agrees with the EQ model, where for the hcp-centered (on-site) structures
the quadrupolar field at the center of the molecules is along $y$, so that the
dipole moment along $z$ is solely due to the small anisotropy in the
polarizability\cite{souza98}. The EQ model also predicts larger 
dipoles in the $C2/m$ than in the $Cmc2_1$ structure
(although the effect is not nearly as pronounced as in the LDA WFs), as well as
an increase in
their magnitude as the molecules move off-site\cite{edwards97,souza98}. 
Although some of the qualitative features of the LDA Wannier dipoles are 
captured by the EQ model, its predictions are
not 
reliable: for instance, it does not reproduce the change in sign of
$q^{\rm s}_z(n)$ for $Cmc2_1$ at low pressures\cite{souza98}; 
other discrepancies can be seen in Table \ref{table_eff_charges}, most notably
in the vibron effective  charges, and are discussed in
Sec.~\ref{dipole_fluct_EQ}. 

\begin{figure}
\centerline{\epsfig{file=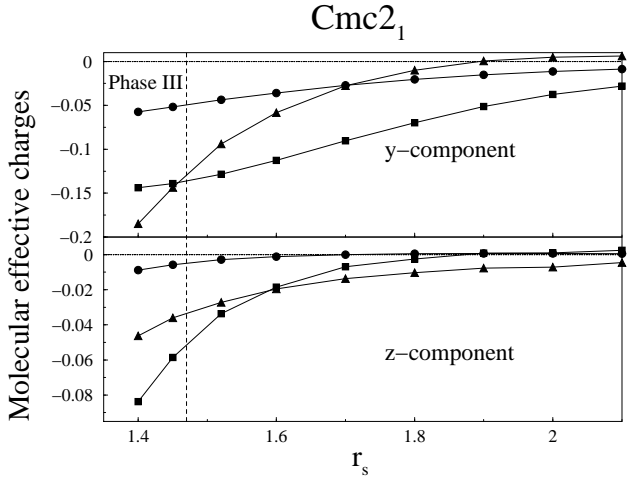,width=3.35in}}
\vspace{0.4cm}
\caption{
Molecular effective charge vectors for molecule 1 in the hcp-centered (on-site)
$Cmc2_1$ structure at several densities. {\Large $\bullet$} Static charge 
(Eq. \ref{static_charge}); $\blacksquare$ dynamical component of effective
charge (last term in 
Eq. \ref{vibron_charge_b}) for the in-phase vibron; $\blacktriangle$ dynamical
component of effective charge for the out-of-phase vibron. The $x$-components 
vanish by symmetry. 
}
\label{fig_charges_cmc21}
\end{figure}

\begin{figure}
\centerline{\epsfig{file=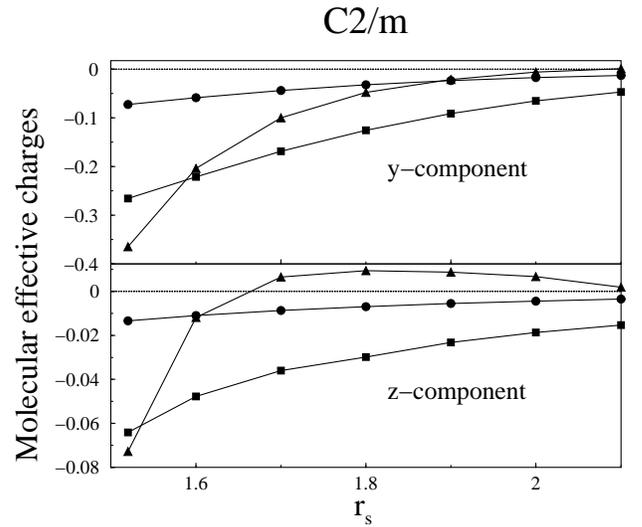,width=3.35in}}
\vspace{0.4cm}
\caption{
Same as Fig. \ref{fig_charges_cmc21}, but for the hcp-centered (on-site) $C2/m$
structure. We have only considered densities for which the LDA band
gap remains open, which are below the density at which
phase III appears. 
}
\label{fig_charges_c2m}
\end{figure}

From Fig. \ref{fig_contour_eq_cmc21} we can already see that 
$|{\bf q}^{\rm s}(n)|$
will be small, since the Wannier distribution is fairly symmetric with respect
to the center of the paired protons. The dependence of the
static charges versus $r_s$ is plotted in Figs. \ref{fig_charges_cmc21} and
\ref{fig_charges_c2m}; as expected they vanish in the low 
density (large $r_s$) limit, and even at the highest
pressures ($\sim~210$~GPa) they are only a few percent of the electron charge,
indicating that, at least in the structures under consideration, 
the ionicity of the molecules remains quite small, contrary to
some proposals\cite{baranowsky92,baranowsky95}. Notice also that at high 
pressures ${\bf q}^{\rm s}(n)$ becomes quite sensitive to the crystal 
structure (see Table \ref{table_eff_charges}).  This is not surprising, since it
is totally induced by the crystal field.
At $\sim 165$~GPa ($r_s=1.45$) the permanent dipole 
moment of an ${\rm H}_2$ Wannier molecule in the hcp-centered $Cmc2_1$ 
structure becomes 0.075~a.u., i.e. more than $1/10$ of the dipole of an isolated water 
molecule ($0.74$~a.u.).

\section{
vibron infrared activity}
\label{dipole_fluct}

\subsection{Wannier-function description}
\label{dipole_fluct_WF}

The investigation of the number of IR-active lattice modes and their oscillator
strengths in candidate structures is a useful
guide in the search for the structures of the high-pressure 
phases\cite{kohanoff99,cui95,souza98}. Here we will focus on the vibron IR
absorption observed in phase III\cite{hanfland93,hemley97}. 
Its relatively strong intensity is somewhat
puzzling, since the stretching mode of the isolated ${\rm H}_2$ molecule is 
Raman-active but IR-forbidden, and this has stimulated a large number of 
studies\cite{kohanoff99,edwards97,hemley97,souza98,baranowsky95,mazin97,hemley94}.

IR absorption is caused by the coupling of light to the change in 
${\bf P}_{\rm mac}$ induced by the lattice modes. 
In the basic theory of IR absorption in molecular 
crystals\cite{califano,schnepp67}, the Clausius-Mossotti approximation 
of nonoverlapping molecules is assumed. The modern theory of 
polarization\cite{ksv} treats rigorously the situation where that 
approximation breaks down, as discussed in the previous section. There, we
decomposed the {\it spontaneous} macroscopic polarization into contributions 
from individual Wannier molecules; here we will do the same for the 
vibron-induced {\it fluctuations} in ${\bf P}_{\rm mac}$.  
For every vibron mode $\nu$ we will assign to each Wannier molecule a 
vibron-induced ``effective charge'' vector (compare with Eq. 
\ref{static_charge}):

\begin{equation}
\label{vibron_charge}
{\bf q}^{\nu}(n) = 
\frac{\partial {\bf d}(n)}{\partial u_{\nu}},
\end{equation}

\noindent where $u_{\nu}$ is the normal coordinate\cite{foot-normal-coord}. The
above expression is very similar to the definition of the Born effective charge
associated with the stretching mode
of a diatomic molecule\cite{ghosez98}. The vector 
${\bf q}^{\nu}(n)$ measures the vibron-induced symmetry-breaking charge
transfer; with the help of Eq. \ref{static_charge}, it can be decomposed into 
two parts (see Eq. 20 of Ref. \cite{ghosez98})\cite{foot-charges}:

\begin{equation}
\label{vibron_charge_b}
{\bf q}^{\nu}(n) = {\bf q}^{\rm s}(n) + u_{\nu} 
\frac{\partial {\bf q}^{\rm s}(n)}{\partial u_{\nu}}.
\end{equation}

\noindent Since ${\bf q}^{\rm s}(n)$ is the static charge, we will call the
second term on the r.h.s. the {\it dynamical} charge.

Experimentally two vibrons have been detected in phase III: 
the lower frequency mode appears in the Raman spectrum, and the 
higher-frequency one in the IR spectrum\cite{mao94,cui95}.
Both the $C2/m$ and the $Cmc2_1$ structures have two vibron modes:
one in which the two molecules in the primitive cell vibrate in-phase
$(\nu={\rm i})$, and a higher frequency mode in which they vibrate out-of-phase
$(\nu={\rm o})$. The effective charges 
${\bf q}^{\nu}(n)$ were calculated using Eq. \ref{vibron_charge} by changing 
the molecular bond length by small amounts $\delta u_{\nu}$ in the range 
$[0.0015,0.0035]$~a.u., after checking that such displacements yield 
essentially linear changes in the Wannier dipoles. Table 
\ref{table_eff_charges} shows their values for $r_s=1.52$; as in the case of
the static charges, we have in general that
$q^{\nu}_{\parallel} > q^{\nu}_{\perp}$.
Figs. \ref{fig_charges_cmc21} and \ref{fig_charges_c2m} plot the 
static and dynamical charges versus $r_s$.
Since these originate from the interactions between molecules, their
magnitudes vanish in the low-pressure (large $r_s$) limit, and increase as 
pressure goes up. The most striking feature is that the dynamical terms 
increase with pressure much more rapidly than the static ones; 
at the highest pressures studied they are already 3.3 to 6.7 times larger,
depending on the structure and on the vibron mode. Since this appears to be a
rather general feature, it is also likely to occur in the 
yet-undetermined structure of phase III; the observed strong vibron IR activity 
is probably caused by this increase of the dynamical charges. 
Their dominant role had been previously inferred 
from the strong anisotropy of the atomic Born effective charge 
tensors\cite{souza98}.

Figs. \ref{fig_charges_cmc21} and \ref{fig_charges_c2m} and Table
\ref{table_eff_charges} also show that the 
vibron-induced fluctuations in the {\it individual} molecular dipoles 
(${\bf q}^{\nu}(n)$), although clearly mode-dependent, are comparable for the 
two vibrons.  (Interestingly, this is not so for the EQ model--see Section
\ref{dipole_fluct_EQ}.)  The important difference occurs only after adding the 
contributions from the two molecules, and can be seen in 
Tables~\ref{table_eff_charges_cmc21} and \ref{table_eff_charges_c2m}.  
In the in-phase mode the large $y$-components 
cancel between the two molecules in the primitive cell, resulting in a weak 
IR activity (which actually vanishes in C2/m, since the small $z$-components 
also cancel).  By contrast, in the out-of-phase mode the large $y$-components 
add up, resulting in a large {\it net} 
$\partial {\bf P}_{\rm mac}/\partial u_{\nu}$, and thus in a strong IR activity. 

\begin{figure}
\centerline{\epsfig{file=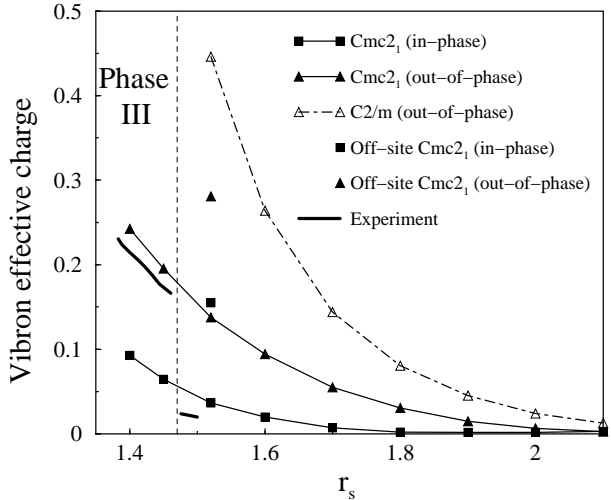,width=3.35in}}
\caption[]{
Magnitude of the net vibron-induced effective charge vector
$|\overline{{\bf q}}^{\nu}|$ (Eq. \ref{net_vibron_charge}) versus $r_s$, for 
the in-phase and out-of-phase vibron modes in 
the $Cmc2_1$ and $C2/m$ structures. Experimental data for phases II and
III is from Ref.~\cite{hemley97}, and was converted from Szigeti to
Born charges~\cite{foot-conversion-szigeti-born}. 
}
\label{fig_ir_activity}
\end{figure}

We emphasize again that all the quantities in 
Eq. \ref{vibron_charge_b} are gauge-dependent, like the Wannier functions
themselves (but see Sec.~\ref{uniqueness}). The gauge-invariant, measurable
quantity is the net ``vibron effective charge'' vector, obtained by averaging 
${\bf q}^{\nu}(n)$ over all molecules in a primitive cell:

\begin{equation}
\label{net_vibron_charge}
{\overline{\bf q}}^{\nu}=\frac{1}{M} \sum_{n=1}^M 
{\bf q}^{\nu}(n) = \frac{1}{n_{\rm mol}} \frac{\partial {\bf P}_{\rm mac}}
{\partial u_{\nu}},
\end{equation}

\noindent where $n_{\rm mol}$ is the number of molecules per unit volume;
the vibron oscillator strength is 
proportional to $n_{\rm mol}{|{\overline{\bf q}}^{\nu}|}^2$. The calculated
values of $|{\overline{\bf q}}^{\nu}|$ versus $r_s$ are plotted in 
Fig.~\ref{fig_ir_activity} together with the experimental results. 
For the out-of-phase mode in the $Cmc2_1$ structure
the LDA calculation yields values very close to the experimentally measured IR
absorption in phase III, but on the other hand the IR activity of the in-phase
mode, although weaker, would still be observed, which is not the case. As for 
$C2/m$, the IR absorption is too strong compared to experiment. Hence it seems
that neither structure is likely to be the correct one for phase III (this is 
also supported by the large number of observed libron modes in phase 
III\cite{goncharov98,hemley98}, which is incompatible with structures with such
small primitive cells). 
Nevertheless, the above results for these structures allow us to make an 
important 
general point.  They show that large permanent molecular dipoles are {\it not} 
required in order for strong vibron IR absorption to occur, contrary to what 
has been
sometimes stated in the literature\cite{edwards97,baranowsky95,pp-edwards}. In
fact, in both
structures the magnitude of the permanent dipoles (static charge) is far too 
small to account by itself for the measured absorption.  However, once the
dynamical charge transfer is accounted for, the resulting IR activity becomes 
even larger than the one measured in phase III.

Finally, Table \ref{table_eff_charges} and Fig.~\ref{fig_ir_activity}
show that the displacement away from the hcp sites significantly
increases the vibron-induced charges, as well as the static charges.
This is yet another example of the strong sensitivity of the effective charges
to the crystal structure, which may help explain the large difference in the 
intensity of IR absorption between phase III and the lower pressure 
phases\cite{souza98,kohanoff99}.

\subsection{Comparison with the EQ model}
\label{dipole_fluct_EQ}

\begin{figure}
\centerline{\epsfig{file=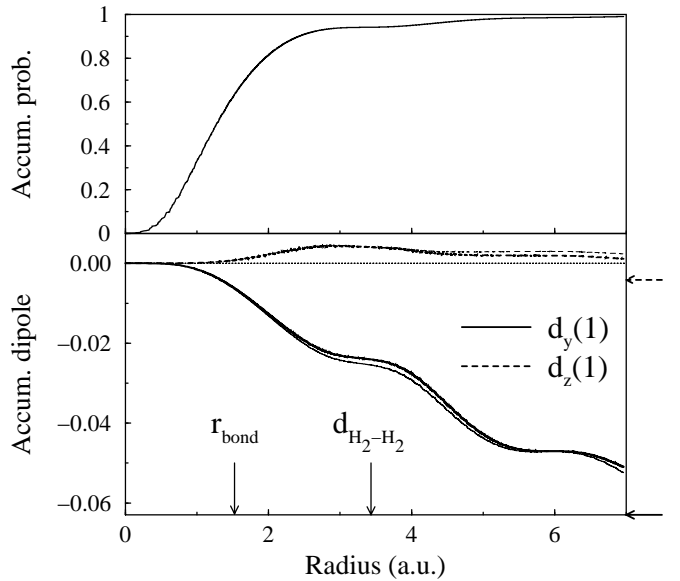,width=3.5in}}
\caption{
Modulus squared (upper panel) and dipole (lower panel) of a WF,
accumulated by integrating up to a
certain radius around the molecular center, for $Cmc2_1$ at $r_s=1.52$.
The $x$-component of the dipole vanishes by symmetry independent of radius, and
the horizontal arrows 
denote the converged values. The thin lines correspond to the WF obtained 
from bond-centered Gaussians with a r.m.s.\ width of 
2.0 \AA\ (see Sec.~\ref{maxloc}). 
}
\label{fig_integrated_charge_dipole}
\end{figure}

Table \ref{table_eff_charges} also contains the values of the vibron-induced
charge vectors on each molecule, as calculated from the EQ model\cite{eq-model}. 
Like the static charges, they are significantly smaller than those obtained from
the WFs. Another important difference is that in the EQ
model the vibron-induced charges on the individual molecules
are more than one order of magnitude smaller 
for the out-of-phase than for the in-phase mode, whereas their WF counterparts
are comparable for the two vibrons. The reason is the following: in the EQ 
model the vibron-induced change in a molecular dipole can be written
to first order as 

\begin{equation}
\label{delta_dipole_EQ}
\delta d_{\parallel} \simeq \delta \alpha_{\parallel}E_{\parallel} +
\alpha_{\parallel} \delta E_{\parallel},
\end{equation}

\noindent where $\alpha$ is the molecular polarizability and $E$ is the 
quadrupolar electric field on the molecular site
(for definiteness we look at the dipole along the 
molecular axis; the same analysis applies to the perpendicular component).
The first term on the r.h.s.
is equal for the two vibrons, so that the difference between their
effective charges arises from the second term. Choosing the 
isolated-molecule parameters from
Refs. \cite{poll78} and \cite{kolos67}, it turns out that the two terms have a 
very similar magnitude. But whereas in the in-phase mode they
have the same sign, in the out-of-phase mode they have opposite signs, so that
their contributions largely cancel, resulting in a much smaller
molecular effective charge. As a consequence, in the $Cmc2_1$ structure the 
in-phase
oscillator strength comes out larger than the out-of-phase, which is the
opposite of the LDA result. 
These discrepancies between the LDA WFs and the EQ model are likely to be 
related at least in part to the 
rather delocalized nature of the induced dipoles, which will be discussed in
the next section, whereas the EQ model assumes point-like (infinitely 
localized) molecules.

\begin{figure}
\centerline{\epsfig{file=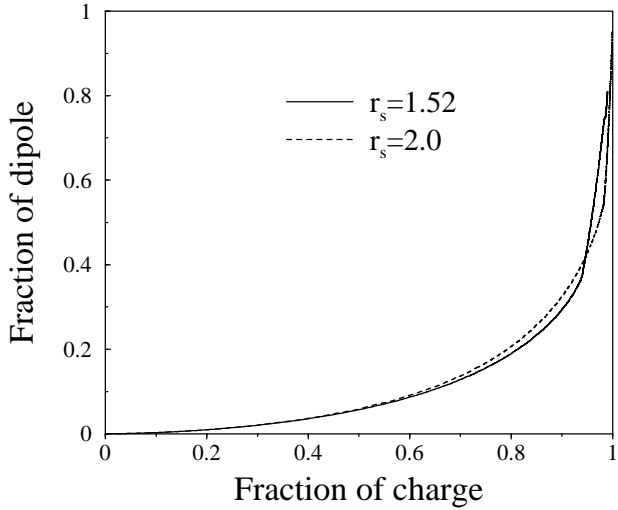,width=3.35in}}
\vspace{0.2cm}
\caption{
Fraction of the Wannier dipole $d_y(1)$ accumulated by integrating up to a
certain radius around the molecular center, versus the fraction of the Wannier
charge that lies inside the same radius, for $Cmc2_1$ (see also 
Fig.~\ref{fig_integrated_charge_dipole}).
}
\label{fig_frac_charge_frac_dipole}
\end{figure}

\section{spatial extent of the Wannier molecules}\label{extent}

In the previous sections we focused our attention on information that can be
extracted from the location of the centers of charge of the WFs.  Here we will 
examine in detail their spatial distribution at high pressure, in particular
their spatial extent. This will allow us to investigate the effects
associated with the overlap between neighboring WFs; 
such effects are expected
to be significant at megabar pressures, as suggested by the large bandwidths, 
in excess of 20~eV (see Fig.~\ref{fig:twobands.ps}).

\subsection{Spread of the Wannier charge and dipole distributions}
\label{spread_charge_dipole}

The spread of the Wannier charge and dipole distributions are presented
in Fig.~\ref{fig_integrated_charge_dipole}. For $Cmc2_1$ at 
$r_s=1.52$ ($r_s=2.0$), a radius of 1.72 a.u.\ (2.25 a.u.), half the shortest 
intermolecular distance $d_{{\rm H}_2-{\rm H}_2}$, encloses about 72\% (85\%)
of the charge and only 17\% (28\%) 
of the $y$-component of the dipole.  This suggests that already at $r_s=2.0$ 
($\sim 13$~GPa) the overlap between nearby WFs is far from negligible. It is 
also clear that the dipole is significantly more spread out than 
the charge, with very large contributions arising from the 
orthogonality tails in the overlap region, where the Wannier charge density is
very small. The longer range of the dipole 
is to be expected, due to the factor ${\bf r}$ in Eq.~\ref{wf_dipole},
but the large magnitude of the effect is somewhat surprising.
Fig.~\ref{fig_frac_charge_frac_dipole} shows even more clearly that for both 
pressures a rather small fraction of the total charge, located in the 
overlapping tails of the WF, is responsible for most of the dipole. Also 
striking is the fact that, at $r_s=1.52$, up to a radius of 7~a.u.\ 
the accumulated $d_z(1)$ remains positive, whereas the converged value
is negative; this suggests that the agreement in sign with the EQ model (see 
Table~\ref{table_eff_charges}) may be
fortuitous, since in that model the dipole is caused by the electric field at 
the center of the molecule. 

\begin{figure}
\centerline{\epsfig{file=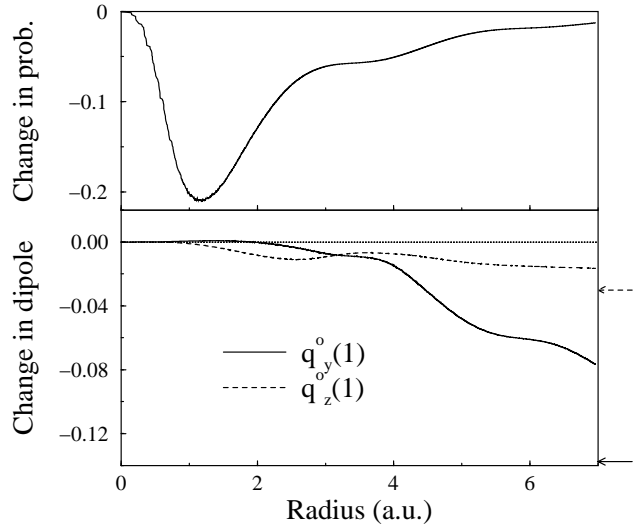,width=3.35in}}
\caption{Upper panel: derivative with respect to the out-of-phase vibron's
normal coordinate $u_{\rm o}$ of the accumulated probability plotted 
in the upper panel of Fig.~\ref{fig_integrated_charge_dipole}. 
Lower panel: derivative with respect to $u_{\rm o}$ of the accumulated dipole 
plotted in
the lower panel of Fig.~\ref{fig_integrated_charge_dipole}, i.e., accumulated
radial
integral of the vibron-induced effective charge 
vector ${\bf q}^{\rm o}(1)$ (Eq. \ref{vibron_charge}) (the $x$-component 
vanishes by symmetry for all radius). The arrows denote 
the converged values.
}
\label{fig_vibron_integrated_charge_dipole}
\end{figure}

By analogy with the radially integrated
Wannier charge and dipole distributions
(Fig.~\ref{fig_integrated_charge_dipole}), one can plot the derivative of these
quantities with respect to the normal coordinate of a vibron mode 
(Fig.~\ref{fig_vibron_integrated_charge_dipole}). 
At high pressures the contributions from the overlapping tails to the change 
in the dipole moment are very significant, even more so than for the 
equilibrium dipole. In other words, in the dense solid the dynamical charge 
transfer processes responsible for the IR activity are very delocalized. 

\begin{figure}
\centerline{\epsfig{file=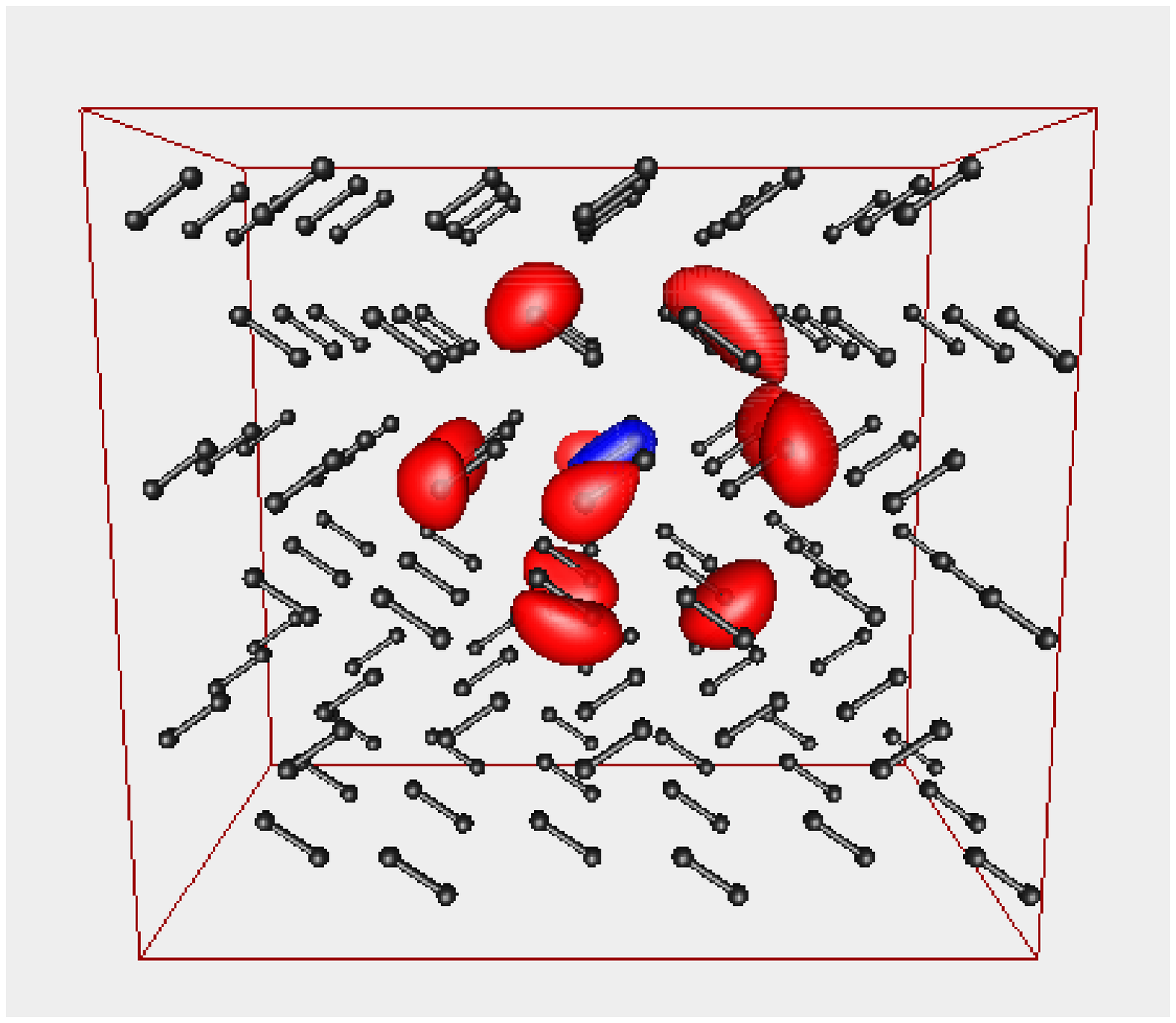,width=3.35in}}
\centerline{\epsfig{file=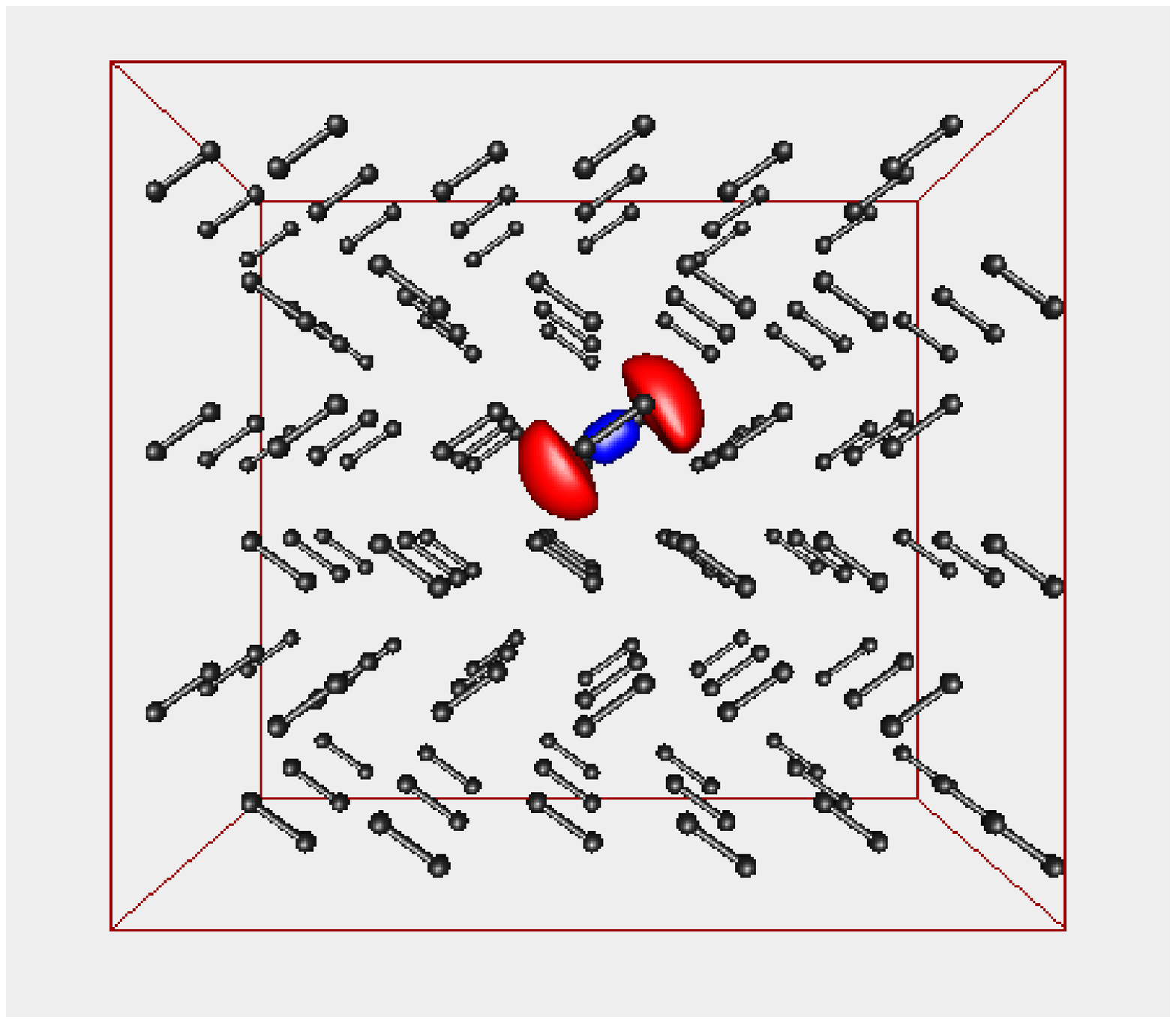,width=3.35in}}
\vspace{0.2cm}
\caption{Upper panel: derivative of the modulus squared of the Wannier orbital
with respect to the normal coordinate of the out-of-phase vibron mode
($v\partial {\left| w_1({\bf r}) \right|}^2 / \partial u_{\rm o}$) for
$Cmc2_1$ at $r_s=1.52$ ($v=59.3$~a.u.\ is the volume of the primitive cell). 
The central contour has an amplitude of $-4.5$, and the
two outer contours have an amplitude of $+0.25$. Lower panel: odd part (with 
respect to the center of the molecule) of the same quantity. The upper (lower)
contour has an amplitude of $+0.05$ ($-0.05$).}
\label{fig_vibron_contour}
\end{figure}

The effect of a vibron on the Wannier charge distribution is depicted in the
upper panel of Fig.~\ref{fig_vibron_contour}: charge is depleted from the inner
part and accumulates in the outer part of the molecule.
Note that the charge transfer occurs mainly along the molecular axis, and is
essentially symmetrical with respect to the molecular center, as one would 
expect from stretching an isolated molecule, which does not break the symmetry
between the two atoms. This kind of charge transfer alone
would lead to a zero net change in the (vanishing) molecular dipole, and hence
to no IR absorption. The contribution of a Wannier molecule to the IR activity
of the crystal comes from the comparatively 
small odd part (with respect to its center) of the charge transfer.
Since this is barely visible in the upper panel of 
Fig.~\ref{fig_vibron_contour}, in the lower panel we have 
removed the large even part. 
It is interesting to note that near the paired protons most of the odd part 
is oriented roughly {\it perpendicularly} to the molecular axis, 
making a
small angle with the $c$ axis, such that
it gives a small positive contribution to $q^{\rm o}_y(1)$ and a larger 
negative contribution to
$q^{\rm o}_z(1)$. This observation is supported by
Fig.~\ref{fig_vibron_integrated_charge_dipole}, which 
shows that for small radius the accumulated $q^{\rm o}_z(1)$ is negative
and larger than the accumulated $q^{\rm o}_y(1)$, which is positive.
For large radius $q^{\rm o}_y(1)$ changes sign 
and ends up overtaking $q^{\rm o}_z(1)$, and the net molecular vibron charge 
vector has a larger projection along the molecular axis than perpendicularly to
it (see Table~\ref{table_eff_charges}).

The results of this section should be relevant for models that attempt to 
account for the
dielectric properties of compressed hydrogen. For instance, it seems unlikely 
that models based on point-like objects, such as the electric quadrupole (EQ)
model\cite{kohanoff99,hemley97,mazin97}, contain all the important ingredients
that lead to the strong IR absorption in the highly compressed phase
III. The reasons are twofold: at such high densities (i) the electrostatic
interactions are expected to differ substantially from the ideal quadrupolar 
one\cite{mazin95} (and in fact the validity of a multipole expansion becomes 
questionable when the molecular charges overlap significantly), and (ii)
the ``classical'' treatment of polarization, based on the bulk $\rho({\bf r})$,
becomes inadequate\cite{ksv}.
We note that although the EQ model can account for both 
static and dynamical charges\cite{hemley97,mazin97,schnepp67}  (see Sec.
\ref{dipole_fluct_EQ}), both effects are then due to {\it local} fields and 
polarizabilities (see also Fig.~2 of Ref. \cite{ghosez98} and associated
discussion regarding local versus nonlocal mechanisms). 

The central conclusion of the
preceding analysis is that the contributions to the induced molecular
dipoles (and their fluctuations)
arising from the overlapping tails of the Wannier orbitals, which extend well 
beyond the nearest neighbor molecules, are crucial. It is instructive
to contrast this state of affairs with what happens in liquid water: there, the
contribution from the orthogonality tails is 
negligible\cite{silvestrelli99b}.
This difference may stem from the fact that an isolated
water molecule is polar, so that the effect of the liquid environment is only
to modify a previously existent dipole moment,
whereas in solid hydrogen the molecular dipole is totally induced.
Induced dipoles tend to be rather extended because the outer regions of the
molecules are most easily polarizable\cite{werner76}; thus
their contribution to the dipole can be large, even though $\rho({\bf r})$ 
is small, because of the ${\bf r}$ factor in 
Eq.~\ref{wf_dipole}. In conclusion, the WF analysis strongly suggests that 
the Clausius-Mossotti picture of nonoverlapping dipoles
breaks down rather dramatically for solid hydrogen at megabar pressures.

\subsection{Measuring the molecular overlap}
\label{measure_overlap}

The overlap between the charge distributions of neighboring WFs can be 
quantified as\cite{silvestrelli99,silvestrelli99b}

\begin{equation}
\label{overlap}
O_{mn} = \frac{ \int {\left| w_m({\bf r}) \right|}^2 
{\left| w_n({\bf r}) \right|}^2 d{\bf r}}  
{{\left( \int {\left| w_m({\bf r}) \right|}^4 d{\bf r}
         \right)}^{1/2}
{\left( \int {\left| w_n({\bf r}) \right|}^4 d{\bf r} 
         \right)}^{1/2}}.
\end{equation}

\noindent For $Cmc2_1$ the largest value of $O_{mn}$
is 0.005 at $r_s=2.0$ and
0.021 at $r_s=1.52$; the latter value is still quite small, roughly twice the
value for WFs located on nearby water molecules in liquid 
water\cite{silvestrelli99b,silvestrelli99}. Thus, by inspection of $O_{mn}$ 
alone one would not suspect that the overlapping tails are so much more 
important for the dipole moments in compressed solid hydrogen than in liquid 
water. The reason is that $O_{mn}$ measures the overlap between 
charge distributions, whereas in this system the dipoles are much
more spread out.

\begin{figure}
\centerline{\epsfig{file=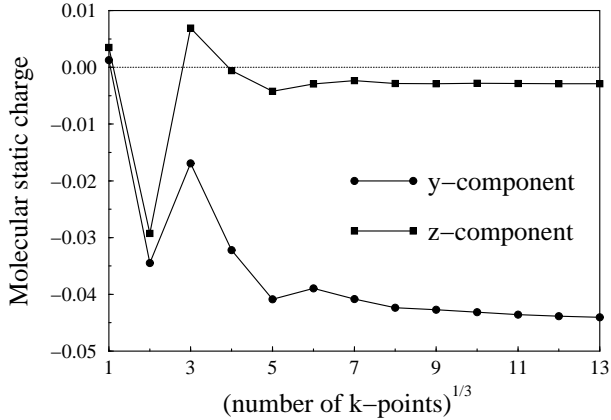,width=3.35in}}
\vspace{0.2cm}
\caption{Static charge vector ${\bf q}^{\rm s}(1)$ (Eq. \ref{static_charge}) 
versus $n$, where $n \times n \times n$ is the size of the mesh of 
$k$-points used for computing the WFs, for the on-site $Cmc2_1$ 
at $r_s=1.52$. The number of $k$-points  is kept fixed at 
$11 \times 11 \times 11$ during the self-consistent calculation.}
\label{fig_dipole_vs_nkpts}
\end{figure}

Another indication that overlap effects are important comes from the well-known
fact that a very large number of $k$-points is required to converge the
total-energy calculations in compressed hydrogen\cite{mazin95}. In fact, if the
molecules were strictly nonoverlapping a single $k$-point would suffice
for computing all physical properties. 
In Fig.~\ref{fig_dipole_vs_nkpts} is shown the static charge (i.e., the dipole)
for WFs obtained using different meshes of $k$-points. It is clear that a
dense mesh is required for converging this quantity.
This results from the fact that,
when using a discrete mesh, the
WFs are actually periodic in real space, with a periodicity which is inversely
proportional to the spacing between neighboring points\cite{marzari97}.  
Hence the need for a fine sampling of the Brillouin zone is just a
manifestation of the large contributions to the Wannier dipole arising from 
the tails far away from the ``home'' unit cell. 
In solid hydrogen a dense mesh of $k$-points is expected to be even more
important for the dielectric properties than for the total energy, since
${\bf P}_{\rm mac}$ is particularly sensitive to the Wannier tails.

\subsection{Quadratic spread and localization length}
\label{second_moment}

\begin{figure}
\centerline{\epsfig{file=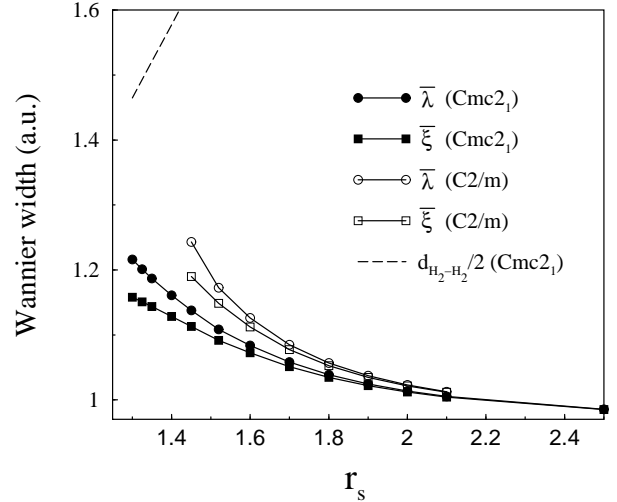,width=3.35in}}
\vspace{0.1cm}
\caption{
Root-mean-square width $\overline{\lambda} = {(\Omega/3M)}^{1/2}$ of the WFs 
and electronic localization length $\overline{\xi}$, both averaged over all 
directions. The dashed line denotes half the shortest intermolecular distance 
in $Cmc2_1$ (in $C2/m$ it is very similar).
}
\label{fig_spread_vs_rs}
\end{figure}

Another way of quantifying the spatial extent of the WFs is in terms of their
quadratic spread. According to Eq. \ref{omega}, the r.m.s.\ width of the Wannier
probability distribution, averaged over all three Cartesian directions and over
the occupied WFs, is $\overline{\lambda} = {(\Omega/3M)}^{1/2}$. This quantity
is plotted versus $r_s$ in Fig.~\ref{fig_spread_vs_rs}.
Notice that it increases with  increasing pressure (decreasing $r_s$), i.e., 
the Wannier molecules become more extended upon compression, 
which is the opposite of what happens in the usual models of molecular
solids\cite{seldam52,keil67}.
This is an overlap effect,
caused by the orthogonality requirement. It is due to the enhancement of the 
outer corona shown in Fig.~\ref{fig_contour_eq_cmc21}, which is not included in
the definition of ``molecules'' in those models. It can also be viewed as
a result of the gap reduction with pressure.

Also plotted in Fig.~\ref{fig_spread_vs_rs} is the electronic localization
length $\overline{\xi} = {(\Omega_{\rm I}/3M)}^{1/2}$\cite{resta99,souza00}, 
where $\Omega_{\rm I}<\Omega$ is the gauge-invariant part of the spread of the
WFs\cite{marzari97}. ${\overline{\xi}}^2$ measures the mean-square quantum
fluctuation of the macroscopic polarization\cite{souza00}, normalized in such a
way as to be finite for insulators, and diverging when the band gap closes.
Notice that in the low-density limit 
$\overline{\lambda} \rightarrow \overline{\xi}$; that happens
because there is only one occupied WF per molecule, and can be understood by 
comparing Eqs. 14 and 15 of Ref. \cite{marzari97}.
As expected, at high pressures $\overline{\lambda}$ and 
$\overline{\xi}$ are larger for the $C2/m$ structure, which has the smaller 
band gap: at $r_s=1.52$ $\overline{\lambda}$ has increased by 13\%
(20\%)
in $Cmc2_1$ ($C2/m$) with respect to the low-density (isolated molecule) value.

\section{Uniqueness of well-localized Wannier molecules}
\label{uniqueness}

As discussed earlier, the WFs are strongly nonunique; in particular, it is only
the sum of all the Wannier dipoles over a primitive cell which is physically
 meaningful (Eq.~\ref{bulk_P}), whereas the individual dipoles (and hence
${\bf q}^{\rm s}(n)$ and ${\bf q}^{\nu}(n)$) are gauge-dependent. Nevertheless,
in this work we have been looking at the individual dipoles of the ``maxloc'' 
WFs in an attempt to extract from them useful physical information. The 
underlying assumption is that in practice well-localized WFs are fairly unique.
For solid hydrogen this is obviously true in the low density limit, where they
reduce to the bonding orbitals of the individual molecules. Here we will 
discuss to what extent that assumption holds for the compressed solid as well.

A systematic way of assessing the degree of uniqueness of well-localized 
WFs would be to implement different localization criteria and then compare
the resulting ``maximally-localized'' WFs. 
We have not attempted such a detailed study; instead we have 
performed a simpler test, which is a first step in that direction. 
As mentioned in Sec.~\ref{maxloc}, in the 
method we are using\cite{marzari97} an initial guess is made for the localized
WFs, with the help of 
``trial functions'', which in our case are bond-centered Gaussians
(we will call the resulting orbitals ``projected WFs''). Their
localization is then enhanced by minimizing the quadratic spread $\Omega$ 
(Eq. \ref{omega}), yielding the ``maxloc'' WFs. 
Since the
projected WFs are totally oblivious of the localization criterion that one
later uses to further localize them, it seems reasonable to assume that
the difference between the projected and the ``maxloc'' WFs is an 
{\it upper bound} to the differences that would occur between WFs obtained 
using any two ``sensible'' localization criteria.

Let us consider the $Cmc2_1$ structure at $r_s=1.52$, for which the average
r.m.s.\ width of the ``maxloc'' WFs is $\overline{\lambda}=1.11$~a.u.
If we choose the r.m.s.\ width of the initial bond-centered Gaussians to be
$1.89$~a.u.\ (1 \AA), the resulting projected WFs are essentially 
indistinguishable from
the ``maxloc'' ones: for instance, the curves corresponding to those in 
Fig.~\ref{fig_integrated_charge_dipole} are virtually identical, and
the individual Wannier dipoles remain the same to at least six significant 
digits! 
This is compelling evidence for a high degree of uniqueness of well-localized 
WFs in this system, at least for the high-symmetry configurations 
that we studied. If we double the width of the initial Gaussian, some 
differences start to appear. They remain barely visible in the accumulated
radial integral of the probability (upper panel of 
Fig.~\ref{fig_integrated_charge_dipole}) but
are noticeable, although still relatively small, in the radially integrated 
dipole (lower panel of Fig.~\ref{fig_integrated_charge_dipole}). For instance,
the large $y$-component of the dipole changes by around 2\%.

\section{Tight-binding analysis}
\label{tb}

In this section, we investigate whether the essential physics of
the WFs in high-pressure H$_2$ phases can be captured by a
simpler tight-binding (TB) approach.  As is well known, the TB
approximation provides a simple, computationally inexpensive
method for computing electronic structure effects, and has the
additional advantage that its output is easily interpreted in
terms of a local, real-space picture\cite{harrison}.  Thus, TB is a natural
approach to explore here, where we want to study the dielectric
properties of H$_2$ phases from just this kind of local point of view.
We confirm below that a previously-proposed $sp^3$ TB model
\cite{chacham:50:94} provides a good description of the occupied
bands in these systems, and show how the operations of constructing
the WFs and computing their contributions to dielectric properties
(such as electric polarization) can be carried out in the TB framework.
Finally, using the fact that the TB representation automatically
provides an atom-by-atom and orbital-by-orbital decomposition, we obtain
useful insights into the nature of the WFs and their contributions
to the dielectric properties.

\subsection{Tight-binding formalism}
\label{tb_formalism}

In the TB method, the Bloch functions $\psi_{n{\bf k}}$
are expanded in a basis of atomic-like orbitals $\phi_{il}$ as
\begin{equation}
\psi_{n{\bf k}}({\bf r})=\sum_{il}C_{n{\bf k}}(il)\,
e^{i{\bf k}\cdot{\bf r}_{il}} \, \phi_{il}({\bf r}) \;\;.
\label{eq:bloch}
\end{equation}
Here $l$ labels the unit cell located at ${\bf R}_l$, $i$ labels
an orbital on the atom at ${\bf r}_{il}={\bf R}_l+\tau_i$ (where
$\tau_i$ specifies the relative position of the atom within the
unit cell), and the vector of coefficients $C_{n{\bf k}}(il)$ forms
the TB representation of the Bloch function.  Our goal is to
carry out a unitary transformation to a set of $M$ localized WFs
\begin{equation}
w_\alpha({\bf r})={1\over N}\,\sum_{n\bf k} U_{\alpha n}^{\bf(k)}\,
\psi_{n{\bf k}}({\bf r})
\label{eq:utran}
\end{equation}
associated with a set of $M$ occupied bands, where the
$U_{\alpha n}^{\bf(k)}$ are $\bf k$-dependent $M\times M$ unitary
matrices that will be fixed by the requirement of maximal
localization \cite{marzari97}.  Introducing the TB representation
of the WF
\begin{equation}
w_\alpha({\bf r})=\sum_{il} W_\alpha(il)\,\phi_{il}({\bf r}) \;\;,
\label{eq:tbwf}
\end{equation}
it follows that
\begin{equation}
W_\alpha(il) ={1\over N}\,\sum_{n\bf k} U_{\alpha n}^{\bf(k)}\,
e^{i{\bf k}\cdot{\bf r}_{il}}\,C_{n\bf k}(il) \;\;.
\label{eq:ctran}
\end{equation}

The essential ingredients needed for the construction of the
maximally-localized WFs \cite{marzari97}, or for the computation of
the Berry-phase polarization \cite{ksv}, are inner products
$\langle u_{n{\bf k}} \vert u_{n'{\bf k'}} \rangle$ between the
cell-periodic part of the Bloch functions
\begin{equation}
u_{n{\bf k}}({\bf r})=e^{-i{\bf k}\cdot{\bf r}}\,\psi_{n\bf k}({\bf r})
\label{eq:udef}
\end{equation}
at nearby $k$-points in the Brillouin zone.  In principle, the
calculation of the $\langle u_{n{\bf k}} \vert u_{n'{\bf k'}} \rangle$
requires a detailed knowledge of the basis orbitals $\phi_{il}$,
which has been done in Ref.~\cite{sanchez-portal00}.
However, in the spirit of minimal empirical TB, we make the approximation
\begin{equation}
\langle u_{n{\bf k}} \vert u_{n'{\bf k'}} \rangle =
\sum_{il} C^*_{n{\bf k}}(il) \, C^{\phantom{*}}_{n'{\bf k'}}(il) \;\;.
\label{eq:tbapprox}
\end{equation}
[When completing the circuit across a Brillouin zone boundary, the
relation $C_{n,{\bf k+G}}(il) = e^{-i{\bf G}\cdot{\bf r}_{il}}\,
C_{n{\bf k}}(il)$ should be used to translate by a reciprocal
lattice vector $\bf G$.]

Eq.~(\ref{eq:tbapprox}) can be derived via a Taylor expansion of the
exponential factor $\exp[i{\bf k}\cdot({\bf r}-{\bf r}_{il})]$, with
the following assumptions:
(i) that the TB basis orbitals are orthonormal,
$\langle \phi_{il} \vert \phi_{i'l'} \rangle =
\delta_{ii'}\, \delta_{ll'}$;
(ii) that the position operator is diagonal in the TB basis,
$\langle \phi_{il} \vert {\bf r} \vert \phi_{i'l'} \rangle =
{\bf r}_{il}\,\delta_{ii'}\, \delta_{ll'}$;
and (iii) that matrix elements of higher powers of the
position operator are likewise trivial,
\begin{equation}
\langle \phi_{il} \vert x^py^qz^r \vert \phi_{i'l'} \rangle =
x_{il}^p \, y_{il}^q \, z_{il}^r 
\,\delta_{ii'}\, \delta_{ll'} \;\;.
\label{eq:higher}
\end{equation}
Conditions (i) and (ii) are actually special cases of (iii) and all
all are quite artificial in that they cannot be satisfied for actual
basis functions.  For
example, while an $sp^3$ hybrid on a given atom should have its charge
center displaced from the geometric center of the atom, condition (ii)
does not allow this effect to be captured.  Similarly, the spread
$\langle \phi_{il}\vert r^2\vert\phi_{il}\rangle
-\langle \phi_{il}\vert {\bf r}\vert\phi_{il}\rangle^2$
of an individual basis orbital is taken to vanish, according to
condition (iii).  Nevertheless, Eq.~(\ref{eq:tbapprox}) is the
logical extension of the empirical TB philosophy, in which one
tries to avoid introducing any additional parameters beyond those
needed to parameterize the Hamiltonian itself.  Despite its
simplicity, the tests presented below demonstrate that this
approach captures much of the interesting complexity of the WFs,
at least for the systems under study here.  A similar
approximation was previously shown to allow for reasonably accurate
TB calculations of dynamical effective charges in semiconductors
\cite{bennetto}.

In practice, we work entirely within the TB representation.
First the $C_{n\bf k}(il)$ are determined on a regular mesh
of $k$-points by solving the standard secular equation involving
the Hamiltonian matrix $H^{(\bf k)}_{ii'}$.  Then the electric
polarization can be computed by inserting Eq.~(\ref{eq:tbapprox})
into the formalism of Ref.~\onlinecite{ksv}.  Similarly,
an ``optimal'' set of unitary matrices $U_{\alpha n}^{\bf(k)}$
can be obtained by inserting Eq.~(\ref{eq:tbapprox})
into the formalism of Ref.~\onlinecite{marzari97}, and from these,
the WFs $W_\alpha(il)$ obtained via Eq.~(\ref{eq:ctran}).
The resulting WFs are optimal in the sense of being maximally
localized in real space, i.e., of minimizing Eq.~(1).

In the ``maxloc'' method \cite{marzari97}, one usually begins by
choosing a set of localized ``trial functions,'' and making a
preliminary unitary rotation among the Bloch orbitals in order
to maximize their projections onto these trial functions,
as discussed in the LDA context at the end of Sec.~\ref{maxloc}.
In the TB context, we have found the following natural way of
constructing the trial Wannier functions.
Since we have two molecules per cell, we want to
carry out the $2 \times 2$ rotation that makes one state have most
of its projection on the first molecule, and the other have most of
its projection on the second molecule.  To do this, we consider the
difference $\Delta \textbf{P} = \textbf{P}_{1} - \textbf{P}_{2}$ of
projection operators $\textbf{P}_{1}$ and $\textbf{P}_{2}$ onto the
first and second molecule, respectively.  (In the TB basis, $\Delta
\bf P$ is just a diagonal matrix with $\pm1$ diagonal entries.)
Then, at each $\bf k$, we diagonalize $\Delta \bf P$ in the space
of the two Bloch states, and set the phase of each eigenvector by
requiring that its inner product with an even linear combination of
$s$ orbitals on the two atoms comprising the molecule should be
real and positive.  We find that the unitary transformation
$\widetilde{U}_{\alpha n}^{\bf(k)}$ obtained in this way turns out
to be an excellent approximation to the ``maxloc''
$U_{\alpha n}^{\bf(k)}$ which minimizes Eq.~(1).  In fact, subsequent
minimization typically only leads to changes of WF coefficients
of order one part in $10^{-5}$, and so that in practice it is not
even necessary to carry out the maxloc minimization procedure.
This is consistent with our similar experience in the LDA context
as discussed at the end of Sec.~\ref{uniqueness}.

\subsection{Details of the tight-binding model}
\label{tb_model}

The TB parameterization we used is the one proposed by Chacham {\em et
al.}~\cite{chacham:50:94}.  These authors showed that the main
characteristics of the electronic structure of high-density solid
hydrogen at megabar pressures could be reproduced by using a
minimal orthogonal TB basis comprised of $s$, $p_x$, $p_y$, and
$p_z$ orbitals on each hydrogen atom.  The intermolecular
matrix elements are taken as
\begin{eqnarray}
V_{ss} &=& V_{ss}(d_{0})e^{\alpha(1-d/d_{0})}
\nonumber \\
V_{pp\sigma} &=& V_{pp\sigma}(d_{0})e^{\beta(1-d/d_{0})}
\nonumber \\
V_{sp} &=& V_{sp}(d_{0})e^{[(\alpha+\beta)/2](1-d/d_{0})}
\nonumber \\
V_{pp\pi}&=&0
\end{eqnarray}
with dimensionless constants $\alpha=5.76$ and $\beta=2.52$, where
$d$ is the interatomic distance and $d_{0}=3.79\,$\AA\ is the
equilibrium hcp lattice constant at zero pressure. Given in Table
\ref{table:tb_parameters} are the $V_{ss}(d_{0})$,
$V_{pp\sigma}(d_{0})$, $V_{sp}(d_{0})$, and the intramolecular
matrix elements $V_{ss}$, $V_{sp}$ and $V_{pp\sigma}$ which are
independent of $d$.  The intra-atomic parameter $\epsilon_{s} -
\epsilon_{p} = -20\,$eV.


\begin{table}
\caption{Tight-binding parameters of
Ref.~\protect\onlinecite{chacham:50:94}, in eV.}
\begin{tabular}{cdd}
\mbox{Hopping parameters} & \mbox{Intramolecular} & \mbox{Intermolecular} \\
\hline
$V_{ss}$        & $-$8.50 & $-$0.04 \\
$V_{sp}$        & $-$8.75 & $-$0.16 \\
$V_{pp \sigma}$ &   +9.00 &   +0.89
\end{tabular}
\label{table:tb_parameters}
\end{table}

In addition to this original tight-binding scheme, we have also
tested an extended scheme that includes a correction designed to
incorporate the effects of the quadrupolar electrostatic fields
arising from neighboring molecules.  In this extended ``TB+Q''
scheme, the molecules are first modeled as point quadrupoles
centered at the molecular sites (mid-bond positions).    The
quadrupole moment tensor for each molecule is taken from the
free-molecule calculations of Ref.~\cite{poll78} by
assuming a linear dependence upon the bond length in the range of
1.4-1.6 a.u.  The total quadrupolar electric field is then
evaluated at each molecular site, and the electrostatic potential
shift on each atom in the molecule is calculated by assuming a
linear extrapolation to the atomic position.  Finally, the
diagonal elements (self-energies) of the TB Hamiltonian matrix
are modified by adding these energy shifts, and the solution
of the secular equation then proceeds as usual.

\subsection{Band structure}
\label{tb_bsr}

We have applied the TB model to the same $Cmc2_{1}$ and $C2/m$
candidate H$_2$ structures studied with LDA methods in earlier
sections, and confirmed that this TB model does a good job of
reproducing the critical features of the electronic band
structure.  For convenience, we present only results on the
$Cmc2_{1}$ geometry; the corresponding results for $C2/m$ are
qualitatively similar.  As already indicated in Sec.~\ref{organization},
our $Cmc2_{1}$ structure has
$r_{s}=1.52$\,a.u., $r_{\rm bond}=1.445$\,a.u, $c/a = 1.576$, and the
tilt angle $\theta=54.0^\circ$. 
The two-molecule (four-atom) unit cell
is illustrated in Fig.~\ref{fig_cmc21}.  Use of the $sp^3$
TB basis leads to a $16 \times 16$ TB Hamiltonian matrix.

\begin{figure}
   \begin{center}
      \epsfig{file=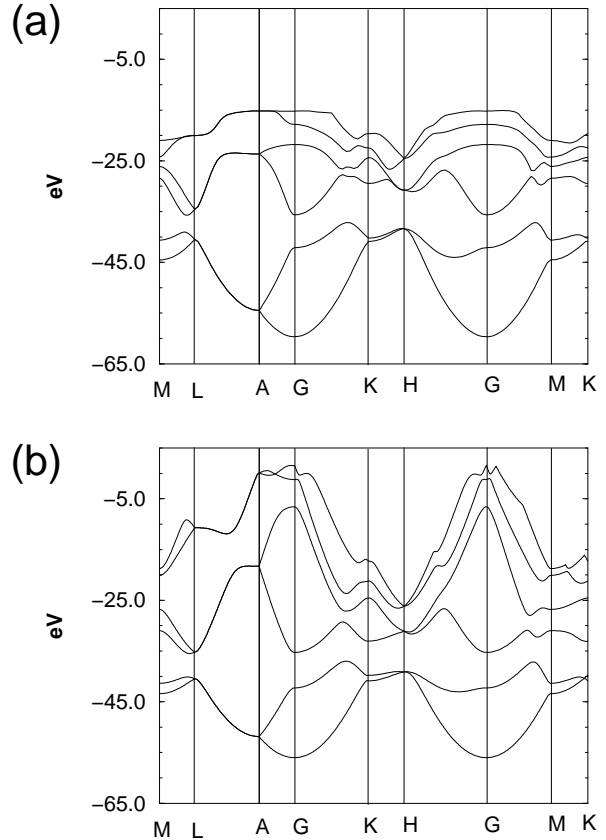,width=3.35in}
   \end{center}
\caption{Electronic band structure calculated by (a)
tight-binding method, and (b) LDA approach.}
\label{fig:twobands.ps}
\end{figure}

Figure \ref{fig:twobands.ps} shows the good agreement between TB
and LDA band structures for this geometry.  The agreement in the
occupied valence-band region (lowest two bands) is excellent, and
the resemblance in the conduction-band region is also reasonable.
The TB model predicts a gap closure at a density of 0.3962
mol/cm$^{3}$, consistent with the results from other
studies~\cite{louie}.
The band structure is hardly affected at all if the TB+Q theory
is used in place of simple TB.

The WFs are constructed for the two occupied bands by using the
definition in Eq.~\ref{eq:ctran}.  A $10 \times 10 \times 10$ $k$-point
mesh is used in calculation. Since there are only two WFs in the unit
cell, and these are related to each other by a symmetry, it suffices to
analyze just one of them.   In Table~\ref{table:wannier_dist} we
analyze the spatial distribution of the WF by decomposing into
contributions coming from the ``home molecule'' and the first two
nearest-neighbor shells of molecules in real space.  The home
molecule is labeled as ``Neighbor 0'', the next six neighboring
molecules form the first shell at a radius of $3.4342$~a.u.\ ($0.9767$
in units of lattice constant), and the second shell comprises the next six
neighboring molecules at a radius of $3.5162$~a.u.\ (one lattice
constant).  Table~\ref{table:wannier_dist} gives a clear picture of the
spatial structure of the WFs.  We can see that the WFs have about 85\%
of their probability on the home molecule, 91\% (cumulatively)
inside the first shell, and 97\% up to the second shell.  In the TB
framework, the contribution to the WFs can be very easily decomposed
further into the $sp_{3}$ tight-binding basis orbitals, as shown in
Fig.~\ref{fig:tb_dist_paper.eps}.  While the $s$-orbital contribution
is $\sim$5 times larger than that of the $p$ orbitals, we find that the
$s$ contribution is almost entirely localized to the home molecule.  On
the other hand, although the $p$ orbitals give a smaller total
contribution, they play a much bigger role in the tail region that
determines the spatial distribution of the WFs.

\subsection{Tight-binding Wannier functions}
\label{tb_wf}

The above quantities have been recomputed using the TB+Q theory
in place of the simple TB theory, but the differences are
not significant.

\begin{table}
\caption{Spatial distribution of Wannier functions for the first three
shells. Shown in the table are the list of the nearest neighboring
molecules, their distance to home unit cell, the contribution to
probability from each molecule, and the accumulated probability up to
the current molecule.}
\begin{tabular}{cccc}
\mbox{Neighbor} & \mbox{Radius~(a.u.)} & \mbox{Probability} & \mbox{Accum. Prob.} \\ \hline

      \mbox{0} & \mbox{0.0000} & \mbox{0.84337} & \mbox{0.84337}  \\ \hline
      \mbox{1} & \mbox{3.4342} & \mbox{0.00764} & \mbox{0.85101}  \\
      \mbox{2} & \mbox{3.4342} & \mbox{0.00764} & \mbox{0.85865}  \\
      \mbox{3} & \mbox{3.4342} & \mbox{0.01294} & \mbox{0.87159}  \\
      \mbox{4} & \mbox{3.4342} & \mbox{0.01294} & \mbox{0.88453}  \\
      \mbox{5} & \mbox{3.4342} & \mbox{0.01898} & \mbox{0.90352}  \\
      \mbox{6} & \mbox{3.4342} & \mbox{0.00648} & \mbox{0.91000}  \\  \hline
      \mbox{7} & \mbox{3.5162} & \mbox{0.00984} & \mbox{0.91984}   \\
      \mbox{8} & \mbox{3.5162} & \mbox{0.00984} & \mbox{0.92969}   \\
      \mbox{9} & \mbox{3.5162} & \mbox{0.01098} & \mbox{0.94067}  \\
      \mbox{10} & \mbox{3.5162} & \mbox{0.01098} & \mbox{0.95165}  \\
      \mbox{11} & \mbox{3.5162} & \mbox{0.00777} & \mbox{0.95942}  \\
      \mbox{12} & \mbox{3.5162} & \mbox{0.00777} & \mbox{0.96719}

\end{tabular}
\label{table:wannier_dist}
\end{table}

According to the modern understanding \cite{ksv}, the electronic
contribution to the polarization can be equivalently expressed either
in terms of a Berry phase (BP) computed from the Bloch functions,
or in terms of the displacements of the Wannier centers (i.e., from
the molecular dipole moments).  These two approaches are compared for
each of the three different computational schemes (TB, TB+Q, and LDA)
in Table~\ref{table:berry_vs_wannier}.  It can be seen that the
bulk polarizations $P_z$ computed from the WF and BP approaches are in
good agreement with each other ($\sim$5\%) for all three schemes,
indicating good internal consistency.  (The small discrepancies can
be traced mainly to incomplete $k$-point convergence.)

\begin{figure}
   \begin{center}
      \epsfig{file=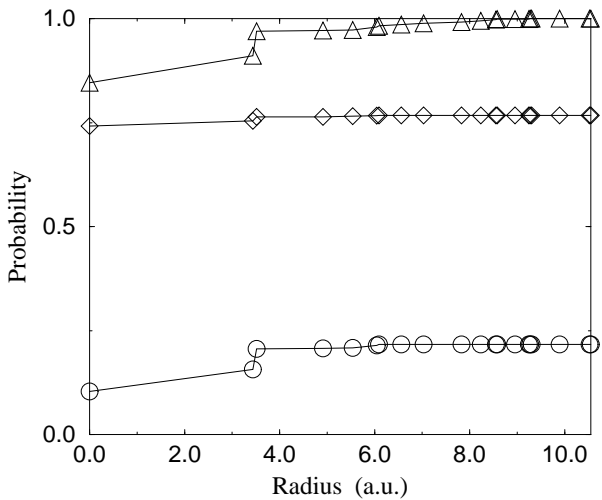,width=3.35in,clip=}
   \end{center}
\caption{The spatial distribution and decomposition of Wannier functions.
$\bigcirc$ is the contribution to probability from $p$-orbitals,
$\Diamond$ is the contribution from the $s$-orbitals, and
($\bigtriangleup$) is the sum of the $s$ and $p$ contributions.}
\label{fig:tb_dist_paper.eps}
\end{figure}

The advantage of the WF approach is that the decomposition into
molecular dipole moments can give some insight into the
microscopic origins of the dielectric properties. 
As in the LDA calculations, we find larger $d_y$ components and
smaller $d_z$ components, with a pattern of signs determined by
symmetry requirements.  The
interpretation of the vibron infrared activities in terms of this
picture has already been discussed in Sec.~\ref{dipole_fluct}.
  
Unfortunately, the level of agreement between the TB and LDA
results for the molecular dipole moments is somewhat
disappointing.  We find a TB $d_y$ value that has the
right sign, and the correct order of magnitude, relative
to the LDA value, but the actual values differ by a factor
of about two.  The TB $d_z$ value even has the wrong sign,
but this is related to the fact that the LDA $d_z$ value happens to
come out very small ($\sim$10 times smaller than for $d_y$;
see also Fig.~\ref{fig_integrated_charge_dipole}).  Thus, it
is not surprising that the {\it relative} TB error in $d_z$ is large,
even though the {\it absolute} TB error is actually smaller for $d_z$
than for $d_y$.  Because the simple TB theory does not
include any charge self-consistency, it was hoped that the
extension to the TB+Q theory might improve the results by
incorporating a leading (quadrupolar) Coulomb contribution.
Table~\ref{table:berry_vs_wannier} shows that the changes from TB
to TB+Q are in the right direction, and there is some
improvement in the $d_z$ (and therefore $P_z$) values, but the
$d_y$ discrepancy is hardly affected.

\begin{table}
\caption{Comparison of dipole moments and bulk polarization
calculated by TB, TB+Q and LDA. Because of the symmetry of $Cmc2_{1}$,
$d_{1x} = d_{2x} = 0$, $d_{1y}=-d_{2y}$ (so $P_{y} = 0 $), 
and $d_{1z}=d_{2z}$. The results from the Wannier-function analysis
are given in the first three rows, while the bulk polarization from
Berry-phase calculation appears in the last row.  Atomic units are used.}
\begin{tabular}{lddd}
& TB & TB+Q & LDA \\
\hline
%
$d_{1y}$     & $-$0.03326 & $-$0.03759 & $-$0.06297 \\
$d_{1z}$     & 0.01252    & 0.00894    & $-$0.00412   \\
$P_z$        & 0.000422    & 0.000301    & $-$0.000139   \\
$P_z$ (BP)   & 0.000436    & 0.000315    & $-$0.000143
\end{tabular} 
\label{table:berry_vs_wannier}
\end{table}

Thus, while the TB theory gives a picture that is
qualitatively correct, it is clear that there is room for
improvement.  Possible avenues for future investigation may
be to consider nonorthonormal TB basis sets, to parameterize and
include off-diagonal matrix elements of the position operators
between TB basis functions, to include $s^*$ or $d$ orbitals
in the TB basis, or to include other Coulomb
contributions (e.g., a self-consistent inclusion of the field
arising from the induced dipoles).

\begin{figure}
   \begin{center}
      \epsfig{file=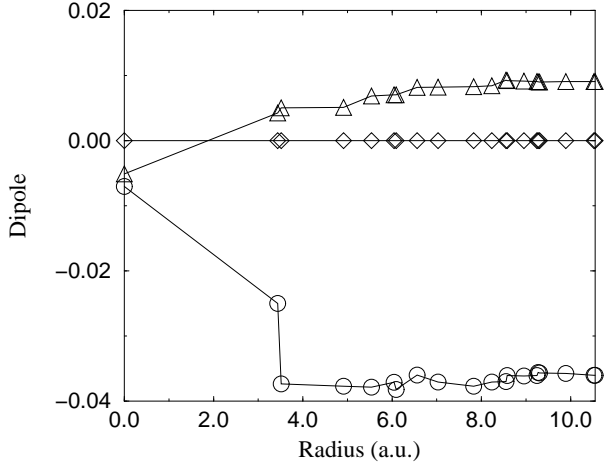,width=3.35in,clip=}
   \end{center}
\caption{Shell-by-shell contribution to dipole. The horizontal line 
( $\Diamond$) is
the dipole along the $x$-direction. $\bigcirc$ and $\bigtriangleup$ indicate
the contributions to dipole along $y$ and $z$ directions,
respectively. The dipole is in a.u.}
\label{fig:tb_dipole.ps}
\end{figure}

Finally, we calculated the shell-by-shell spatial decomposition of
the molecular dipole moments in the TB scheme.  The results appear
in Fig.~\ref{fig:tb_dipole.ps}.  (By symmetry, all $d_x$
contributions are identically zero.) Comparing
Fig.~\ref{fig:tb_dipole.ps} with Fig.~\ref{fig:tb_dist_paper.eps},
one sees that although the contribution up to the first neighbor shell
is 91\% for the density, it is only 62\% and 60\% for $d_{y}$ and
$d_{z}$, respectively.  Up to the second shell, the contribution is
97\% for the density and 95\% for $d_{y}$, but only 67\% for $d_{z}$.
(Of course, $d_{z}$ is the only component that survives in
the summation giving the bulk polarization.)  Thus, the dipoles are
found to be very delocalized, spanning over quite a few neighboring
molecules, in agreement with the LDA results.

To summarize this section, we have demonstrated that an empirical TB
framework allows for a very useful qualitative (and often
semiquantitative) analysis of the WFs in systems such as the H$_2$
phases under study here.  It is typical of the empirical TB approach
that one cannot insist on quantitative accuracy at the level of
first-principles schemes.  However, the TB approximation has proven
enormously useful over the years because of its simplicity,
transparency, and ease of application.  These features often allow for
insightful modeling of simple systems, or for efficient calculations of
large and complex systems where {\it ab-initio} schemes would not be
practical.  For example, one could easily use the present scheme for a
computationally efficient analysis of the local dielectric structure of
more complex H$_2$ crystal structures \cite{kohanoff99} or of supercell
realizations of disordered H$_2$ systems\cite{chacham:50:94}.  We
expect that the coupling of Wannier and TB approaches will prove to be
a useful strategy in a wide variety of other materials systems as well.

\section{discussion}
\label{discussion}

\subsection{Wannier functions and the Clausius-Mossotti approximation}

As discussed by Harrison\cite{harrison}, WFs provide a rigorous formulation of
the ``extended bond orbitals'' which appear in tight-binding models. It should
be noted that orbitals more localized than WFs can be constructed by removing 
the orthogonality constraint, which gives rise to the long-range tails; the
resulting orbitals correspond to the ``bond orbitals'' in Harrison's picture.
However, the straightforward connection to the polarization is then lost, since
${\bf P}_{\rm mac}$ is no longer simply the sum of the dipole
moments of those orbitals, and additional cross-terms connecting 
orbitals in different cells appear. Therefore, as far as dielectric properties
are concerned, the ``maxloc'' WFs seem to provided the most
localized description of the electronic structure.  In particular, the spatial 
extent of their dipole moments 
(Figs.~\ref{fig_integrated_charge_dipole} and \ref{fig:tb_dipole.ps})
provides a natural length scale to compare with
the intermolecular distance in order to assess the validity of the
Clausius-Mossotti approximation of nonoverlapping dipoles. 

\subsection{Identifying ``molecules'' in the dense solid}
\label{mol_in_solid}

There is a vast literature dealing with useful ways of
identifying individual ``atoms'' inside a molecule, or individual
``molecules'' in a dense medium (see Refs.~\cite{bader90,site99,batista99} and
references cited therein). In this paper we have advocated
using ``maxloc'' WFs as a useful computational definition of 
``${\rm H}_2$ molecules'' in the insulating molecular solid 
(similarly, in
Refs.~\cite{silvestrelli99b,silvestrelli99} WFs were used for defining
``water molecules'' in liquid water).
In spite of having some counterintuitive features (becoming larger under
pressure) and some conceptual limitations (being defined only in the 
independent-electron framework; not being totally unique, since they depend on
the measure of localization), 
they have the following important conceptual advantages.
(i) When the Clausius-Mossotti approximation breaks down -~which is
when the ambiguity in identifying individual ``molecules'' appears~- the
sum of the dipoles of the Wannier molecules still gives the bulk polarization
exactly (Eq. \ref{bulk_P}).  By contrast, {\it any} definition of ``molecules''
based upon a direct 
partition of the bulk electronic charge density necessarily yields an 
{\it incorrect} result, since away from the Clausius-Mossotti limit the
information about the bulk ${\bf P}_{\rm mac}$ is not in 
$\rho({\bf r})$\cite{ksv}. For example, in Ref.~\cite{batista99} it was found
that different schemes for partitioning the charge density yield very different
molecular dipoles in liquid water. We would expect the situation with such
approaches to be
even more severe in the case of solid hydrogen, since the dipoles are smaller
and originate mostly in the Wannier tails.
(ii) At high densities the WFs interpenetrate one another, so that 
the charge density at a given point is a sum of
contributions from different molecules.  Thus, effects related to molecular 
overlap are naturally discussed in the Wannier representation.

\subsection{Intramolecular versus intermolecular charge transfer}
\label{wf_tb}

There has been some debate about whether the strong vibron IR activity in
compressed solid hydrogen
is due mainly to intramolecular or intermolecular charge 
transfer\cite{hemley97,baranowsky95,hemley94}.
As a result of the ambiguity in defining ``molecules'' in the solid,
the question is to some extent ill-posed. 
Of course, there is no charge transfer between 
``Wannier molecules'', since their charge is fixed. 
However, a heuristic argument can
be attempted: the fraction of the vibron-induced molecular effective charge
originating in
the central part of the WF can be viewed as the ``intramolecular'' contribution
(polarization of the molecular bond), whereas the contributions from the 
orthogonality tails in the overlap regions are of
``intermolecular'' origin.
As the pressure increases, the 
``intramolecular'' part of the WF becomes smaller and contains less charge
(and presumably becomes less polarizable),
whereas the opposite happens to the outer corona. The results of
Secs.~\ref{extent} and \ref{tb} suggest that as far as 
polarization--and hence IR activity--are concerned, at megabar pressures
the ``intermolecular'' contribution associated with the outer corona 
is dominant.

\subsection{Summary}
\label{summary}

Using a method for computing well-localized Wannier functions\cite{marzari97}, 
we have 
presented a ``chemical-like'' localized picture of the electronic structure of
solid molecular hydrogen, and used it to investigate the dielectric properties
of the
compressed system. This approach is particularly well-suited for studying the
effects of molecular overlap, which become increasingly more 
important at high pressures.
We found the somewhat surprising result that already at moderate pressures the
orthogonality tails of the WFs in the overlap regions give rise to most of the
induced dipole moments on the ``Wannier molecules''; this clearly indicates a 
breakdown of the Clausius-Mossotti approximation. Under those circumstances
the electric polarization cannot be extracted from the electronic charge 
density in the unit cell, and the Berry-phase/Wannier-function theory\cite{ksv}
must be used instead. 
The present approach clarifies the origin of the strong
vibron IR activity in phase III and identifies the dominant mechanism: 
even though the 
{\it permanent} dipoles of the molecules in our prototype structures
are too small to account for the vibron oscillator strength, the 
vibron-induced dipole {\it fluctuations} are of the 
right order of magnitude in $Cmc2_1$, and actually too large in $C2/m$. 
In other words, in the strongly compressed solid
the dynamical contribution to the vibron effective
charge dominates the static one. This conclusion seems to be supported by the
fact that, even though several libron modes have been identified in phase
III\cite{goncharov98,hemley98}, no strong libron IR activity has been
reported. If the molecules had any significant spontaneous 
polarization, it should manifest itself in the librational IR absorption.
Thus, we see that well-localized Wannier functions provide a useful
definition of ``${\rm H}_2$ molecules'' in the dense solid, which can be
used to gain important insight into
the microscopic mechanisms of its dielectric response.

\section*{Acknowledgments}

I.S. and R.M.M. acknowledge financial support from DOE Grant 
No. DEFG02-96-ER45439, X.Z. and D.V. from NSF Grant DMR-9981193.
I.S. acknowledges financial support from FCT (Portugal).


\begin{references}
\bibitem{foot-cmca} A class of orthorhombic structures, with the $Cmca$ 
symmetry, was also found to be energetically 
competitive\cite{edwards96,kitamura00}. However,
these structures are metallic, in contradiction with experiment.
\bibitem{eq-model} The ${\rm H}_2$ molecules are modeled as point objects with
a cylindrically symmetric quadrupole moment and polarizability, which become 
polarized in the quadrupolar electric field of the surrounding molecules. The 
values of the quadrupole moment and polarizability as a function of bond length
are from Refs. \cite{poll78} and \cite{kolos67} respectively.
\bibitem{foot-normal-coord} Since we are neglecting the small coupling between
the vibron and the other lattice modes, we choose the normal coordinate to be 
the change in molecular bond length $r_{\rm bond}(1)$ of molecule 1 from its 
equilibrium value: $u_{\nu} = \delta r_{\rm bond}(1)$.
\bibitem{foot-charges} In Ref. \cite{ghosez98}
Eq. \ref{vibron_charge_b} is discussed for an isolated diatomic 
molecule. In that case the effective charge vector is directed along the 
molecular axis, so that a scalar definition is adequate, since at most the 
signs of the static and dynamical terms may differ. The 
vectorial definition becomes relevant for
the Wannier molecules in the solid: because there the symmetry is lower, the 
effective charge vectors can point along other directions, which in 
general also differ between the static and dynamical terms.
\bibitem{foot-conversion-szigeti-born} The results in Ref. \cite{hemley97}
are presented in terms of Szigeti effective charges, obtained, under the
approximation of cubic symmetry, by
multiplying the Born charges by $[ \epsilon(\infty)+2]/3$\cite{ghosez98}.
We converted them back to Born charges using
$\epsilon(\infty) = n^2=(2.7)^2$\cite{hemley96}.
Moreover, we multiplied the effective charges given in 
Ref. \cite{hemley97} by
$\sqrt{3}$, in order to take into account the averaging over all directions,
due to the use of polycrystalline samples. Unlike the Szigeti charges,
which are model-dependent quantities since they are related to local fields,
the Born charges are macroscopic, model-independent, measurable quantities.
\end{references}
\end{document}